\documentclass[ onecolumn, letterpaper]{article}
\newtheorem{theorem}{Theorem}
\newtheorem{exmpl}{\mdseries\itshape Example}
\newenvironment{ex}{\begin{exmpl}\upshape}{\end{exmpl}}
\newtheorem{lemma}{Lemma}
\newtheorem{defn}{Definition}
\newenvironment{definition}{\begin{defn}\upshape}{\end{defn}}
\usepackage{graphicx}
\usepackage{fancyvrb}
\usepackage{latexsym}

\title{ Completeness and Decidability Properties for Functional Dependencies in
XML
}
\author{ Millist  W. Vincent  and Jixue Liu \\
\\
\and Advanced Computing Research Centre,
School of Computer and Information Science\\
\and  The University of South Australia
\and The Levels, SA5095, Adelaide, Australia\\
\and Email: millist.vincent @unisa.edu.au}

\begin{document}
\date{}
\maketitle

\begin{abstract}
XML is of great importance in information storage and retrieval because of its
recent emergence as a standard for  data representation and interchange on the
Internet. However XML provides
little semantic content and as a result several papers have addressed  the
topic of
how to improve the semantic expressiveness of XML. Among the most important of
these   approaches has been that of defining integrity constraints in XML. In a
companion paper we defined strong functional dependencies in XML(XFDs).  We
also presented a set of axioms for reasoning about the implication of XFDs and
showed that the axiom system is sound for arbitrary XFDs. In this paper we
prove
that the axioms are also complete for unary XFDs (XFDs with a single path
on the
l.h.s.).  The second contribution of the paper is to prove that the implication
problem for unary XFDs is decidable and to provide a linear time algorithm
for it.

\end{abstract}
\section{Introduction}
The eXtensible Markup Language (XML) \cite{Bra98} has recently emerged as a
standard for data representation and interchange on
the Internet \cite{Wid99, Abi00}.  While providing syntactic flexibility, XML
provides
little semantic content and as a result several papers have addressed  the
topic of
how
to improve the semantic expressiveness of XML.  Among the most important of
these   approaches has been that of defining integrity constraints in XML
\cite{Bun01a, Fan01a}.  Several different classes
of integrity constraints for XML have been defined including key constraints
\cite{Bun01, Bun01a, Bun02}, path constraints \cite{Abi97, Bun99, Bun01a,
Bun98},
and
inclusion constraints \cite{Fan00, Fan01} and properties such as
axiomatization and
satisfiability have been investigated for these constraints.  One
observation to make
on this research is that the flexible structure of XML makes the
investigation of
integrity constraints in XML more complex and subtle than in relational
databases.  However, one topic that has been
identified as an open problem in XML research \cite{Wid99} and which
has been little investigated is how to extended the oldest and most well
studied
integrity constraint in relational databases, namely {\em functional
dependencies}
(FDs),
to XML and then how to develop a normalization theory for XML. This problem is
not of just theoretical interest. The theory of FDs and normalization forms the
cornerstone of practical relational database design  and the development of
a similar
theory for XML will similarly lay the foundation for understanding how to
design
XML documents. In addition, the study of FDs  in XML is important
because of the close connection between XML and relational databases. With
current technology, the source of XML data is typically a relational database
\cite{Abi00} and relational databases are also normally used to store XML data
\cite{Sha99}. Hence, given that FDs are the most important constraint
in relational databases,
the study of FDs in XML assumes heightened importance over other
types of constraints which are unique to XML \cite{Bun01b}.
The only papers
that have specifically  addressed this problem are the recent papers
\cite{Are02,Vin02a}.   Before presenting  the contributions of \cite{Are02,
Vin02a}, we  briefly outline the approaches to defining FD satisfaction in
incomplete relational databases.

There are two approaches, the first called the {\em weak satisfaction}
approach and
the other called the {\em strong satisfaction} approach \cite{Atz93}.  In
the weak
satisfaction approach, a relation is defined to weakly satisfy a FD if
there exists
{\em at least one} completion of the relation, obtained by replacing all
occurrences
of nulls by data values, which satisfies the FD.  A relation is said to
strongly satisfy
a FD if {\em every} completion of the relation satisfies the FD.  Both
approaches
have their advantages and disadvantages (a more complete discussion of this
issue
can be found in \cite{Vin02a}).  The weak satisfaction
approach has the advantage of allowing a high degree of uncertainty to be
represented in a database  but at the expense of making maintenance of
integrity
constraints much more difficult. In contrast, the strong satisfaction approach
restricts the amount of uncertainty  that can be represented in a database
but makes
the maintenance of integrity constraints much easier. However, as argued in
\cite{Lev99}, both approaches have their place in real world applications and
should be viewed as complementary rather than competing approaches.  Also,
it is
possible to combine the two approaches by having some FDs in a relation
strongly
satisfied and others weakly satisfied \cite{Lev98b}.

The contribution of \cite{Are02} was, for the first time, to define FDs in XML
(what we call XFDs) and then to define a normal form for a XML document based
on the definition of a XFD.  However, there are  some difficulties with the
definition of a XFD given in \cite{Are02}.  The most fundamental problem is
that
although it is
explicitly recognized in the definitions that XML documents have missing
information, the definitions in \cite{Are02}, while having some elements of the
weak instance approach, are not a strict extension of this approach since
there are
XFDs that are violated according to the definition in \cite{Are02} yet
there are
completions of the tree that satisfy the XFDs (see
\cite{Vin02a} for an example).  As a result of this it is not clear that
there is any
correspondence between FDs in relations and XFDs in XML documents. The other
difficulty is that the approach to defining XFDs is not straightforward and
is based
on the complex and non-intuitive notion of a "tree tuple".

In \cite{Vin02a} a different approach was taken to defining XFDs which
overcomes
the difficulties just discussed with the approach adopted in \cite{Are02}. The
definition in \cite{Vin02a} is based on extending the strong satisfaction
approach to
XML.  The definition of a XFD given in \cite{Vin02a} was justified formally by
two main results.  The first result showed that for a very general class of
mappings
from an incomplete relation into a XML document, a relation strongly
satisfies a
unary FD (only one attribute on the l.h.s. of the FD) if and only if the
corresponding XML document strongly satisfies the corresponding XFD.  The
second result showed that a XML document strongly satisfies a XFD if and
only if
every completion of the XML document also satisfies the XFD. The other
contributions in \cite{Vin02a} were firstly to define a set of axioms for
reasoning
about the implication of XFDs and show that the axioms are sound for arbitrary
XFDs.  The final contribution was to define a normal form, based on a
modification of the one proposed in \cite{Are02}, and prove that it is a
necessary
and sufficient condition for the elimination of redundancy in a XML document.

The contribution of this paper is to extend the work in \cite{Vin02a} in two
important ways.  As just mentioned, in \cite{Vin02a} a set of axioms for XFDs
were provided and shown to be sound.  In this paper we prove that the
axioms are
also complete for unary XFDs.  The second contribution of the paper is to prove
that the implication problem for unary XFDs is decidable and to provide a
linear
time algorithm for it.  These results have considerable significance in the
development of a theory of normalization for XML documents.  In relational
databases, the classic results on soundness and completeness of Armstrong's
axioms
\cite{Arm74} and the resulting closure algorithm for FD implication play an
essential role in determining whether a relation is in one of the classic
normal
forms.  Similarly, the results in this paper are an important first step in the
development of algorithms for testing the normal form proposed in
\cite{Vin02a}.
In addition, the result on completeness is of theoretical interest in
itself since it
ensures that there are no other 'hidden' axioms for reasoning about the
implication
of XFDs.

The rest of this paper is organized as follows. Section 2 contains some
preliminary
definitions.  In Section 3 a XFD is defined.  In Section 4 axioms for XFDs are
presented and are shown to be sound for arbitrary XFDs and complete for unary
XFDs.  In Section 5 the implication problem for unary XFDs is investigated
and a
linear time algorithm for the implication problem is presented and shown to be
correct.  Finally, Section 6 contains concluding comments.

\section{Preliminary definitions}
In this section we present some preliminary definitions that we need before
defining
XFDs.
We firstly present the definition of a XML tree adapted from the definition
given in
\cite{Bun01a}.

\begin{definition}
\label{D:treedef}
Assume a countably infinite set ${\bf E}$ of element labels (tags), a
countable infinite set ${\bf A}$ of attribute names and a symbol
\texttt{S} indicating text. An {\em XML tree} is  defined to be $T =
(V, lab, ele, att,val, v_r)$ where $V$ is a {\em finite}  set of nodes  in $T$;
$lab$ is a function from $V$ to ${\bf E} \cup {\bf A} \cup
\texttt{\{S\}}$; $ele$ is a partial function from $V$ to a sequence of $V$
nodes such that for any $v \in V$, if $ele(v)$ is defined then
$lab(v) \in {\bf E}$; $att$ is a partial function from $V \times {\bf
A}$ to $V$ such that for any $v \in V$ and $l \in {\bf A}$, if $att(v,
l) = v_1$ then $lab(v) \in {\bf E}$ and $lab(v_1) = l$; $val$ is a
function such that for any node in $v \in V, val(v) = v$  if $lab(v) \in
{\bf E}$ and $val(v)$ is a string if either $lab(v)$ = \texttt{S} or
$lab(v) \in {\bf A}$; $v_r$ is a distinguished node in $V$ called the
$root$ of $T$ and we define $lab(v_r) = root$. Since node identifiers are
unique, a
consequence of the definition of $val$ is that if $v_1 \in {\bf E}$ and
$v_2 \in {\bf
E}$ and $v_1 \neq v_2$ then $val(v_1) \neq val(v_2)$. We also extend the
definition of $val$ to sets of nodes and if $V_1 \subseteq V$, then
$val(V_1)$ is the
set defined by $val(V_1) = \{val(v) | v \in V_1\}$.

For any $v \in V$, if $ele(v)$ is defined then the nodes in $ele(v)$ are
called {\em subelements} of $v$.  For any $l \in {\bf A}$, if $att(v, l)
= v_1$ then $v_1$ is called an {\em attribute} of $v$. Note that a XML tree $T$
must be a tree.  Since $T$ is a tree the ancestors of a node $v$,
denote by $Ancestor(v)$ are defined as in Definition  ~\ref{D:treedef}. The
children of a node $v$ are also defined as in Definition ~\ref{D:treedef}
and we
denote the parent of a node $v$ by $Parent(v)$.

\end{definition}
We note that our definition of $val$ definition differs slightly from that in
\cite{Bun01a} since we have
extended the definition of the $val$ function so that it is also defined on
element
nodes.   The reason for this is
that we want to include in our definition paths that do not end at leaf
nodes, and when we do this we want to compare element nodes by node identity,
i.e. node equality, but when we compare attribute or  text nodes we want to
compare them by their contents, i.e. value equality. This point will become
clearer
in the
examples and definitions that follow.

\begin{figure}[here]
\begin{center}
\includegraphics[scale=.6]{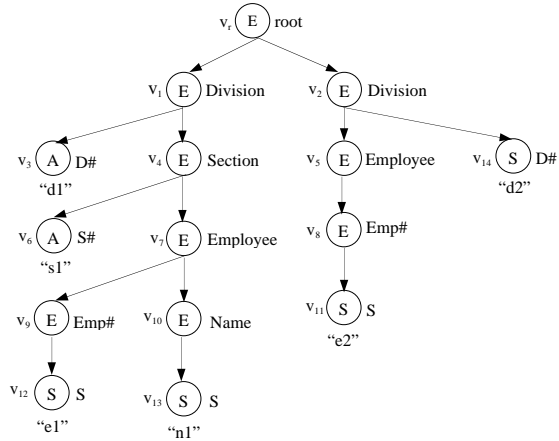}
\end{center}
\caption{A XML tree }
\label{F:Xtree}

\end{figure}

We now give some preliminary definitions related to paths.
\begin{definition}

A  {\em path} is an expression of the
form   $l_1.\cdots.l_n$, $n \ge 1$, where $l_i \in
{\bf E} \cup {\bf A} \cup \texttt{\{S\}}$ for all $i, 1 \leq i \leq n$ and
$l_1 =
root$.
If $p$ is the path  $l_1.\cdots.l_n$ then $Last(p)$ is $l_n$.
\end{definition}

For instance, in Figure ~\ref{F:Xtree},  \texttt{root} and
\texttt{root.Division} are
paths.

\begin{definition}
Let $p$ denote the path $l_1.\cdots.l_n$. The function $Parnt(p)$ is the path
$l_1.\cdots .l_{n-1}$.  Let $p$ denote the path $l_1.\cdots.l_n$ and let $q$
denote the path $q_1.\cdots .q_m$.  The path $p$ is said to be a {\em
prefix} of the
path $q$ if
$n \leq m$ and $l_1 = q_1, \ldots, l_n = q_n$. Two paths $p$ and $q$ are equal,
denoted by $p=q$, if $p$ is a prefix of $q$ and $q$ is a prefix of $p$. The
path
$p$ is said to be a {\em strict
prefix} of $q$ if $p$ is a prefix of $q$ and $p \neq q$. We also  define the
intersection of two paths $p_1$ and
$p_2$, denoted but $p_1 \cap p_2$, to be the maximal common prefix  of
both paths. It is clear that the
intersection of two paths is also a path.
\end{definition}

For example, in Figure ~\ref{F:Xtree}, \texttt{ root.Division} is a strict
prefix of
\texttt{root.Division.Section } and \texttt{ root.Division.d\#} $\cap$
\texttt{root.Division.Employee.Emp\#.S} = \texttt{ root.Division}.

\begin{definition}
\label{D:pathdef}
A {\em path instance} in a XML tree $T$
is a sequence $\bar{v}_1. \cdots. \bar{v}_n$  such that $\bar{v}_1 = v_r$
and for
all $\bar{v}_i, 1 < i
\leq n$,$v_i \in V$ and $\bar{v}_i$ is a child of $\bar{v}_{i-1}$.   A path
instance
$\bar{v}_1. \cdots. \bar{v}_n$ is said to be {\em defined over the path}
$l_1.\cdots.l_n$ if for all $\bar{v}_i, 1 \leq i
\leq n$, $lab(\bar{v}_i) = l_i$.
Two path instances $\bar{v}_1. \cdots. \bar{v}_n$ and
$\bar{v}'_1. \cdots. \bar{v}'_n$ are said to be {\em distinct} if $v_i \neq
v'_i$ for
some $i$, $1
\leq i \leq n$. The
set of path instances over a path $p$ in a tree $T$ is denoted by $Paths(p)$
\end{definition}

\begin{definition}
An {\em extended XML tree} is a tree $(V \cup {\bf N}, lab, ele, att,val,
v_r)$
where ${\bf N}$ is a set of marked nulls that is disjoint from $V$ and if
$v \in {\bf
N}$ and
$v \notin {\bf E}$ then $val(v)$ is undefined.
\end{definition}

\begin{definition}
Let $T$ be a XML tree and let $P$ be a set of paths.   Then $(T, P)$ is {\em
consistent} if:

(i)  For any two
paths $l_1.\cdots.l_n$ and $l'_1.\cdots.l'_m$ in $P$ such that  $l'_m =
l_i$ for
some $i$, $1 \leq i \leq n$ then  $l_1.\cdots.l_i = l'_1.\cdots.l'_m$;

(ii) If $v_1$ and $v_2$ are two nodes in $T$ such that $v_1$ is the parent of
$v_2$, then there exists a path $l_1.\cdots.l_n$ in $P$ such that there
exists $i$ and
$j$, where $1 \leq i \leq n$ and $1 \leq j \leq n$ and $i < j$ and
$label(v_1) = l_i$
and $label(v_2) = l_j$.
\end{definition}

\begin{definition}
Let $T$ be a XML tree and let $P$ be a set of paths and such that $(T, P)$ is
consistent.  Then  a minimal extension of $T$, denoted by
$T_P$, is an extended XML tree constructed as follows.  Initially let
$T_P$ be $T$.  Process each path $p$  in $P$ in an arbitrary order as
follows. For
every node in $v$ in $T$  such that $lab(v)$ appears
in $p$ and there does not exist a path instance containing $v$ which is
defined over
$p$, construct a path instance over $p$ by adding
nodes from ${\bf N}$ as ancestors and descendants of $v$.
\end{definition}

The next lemma follows easily from the construction procedure.

\begin{lemma}
$T_P$ is unique up to the labelling of the null nodes.
\end{lemma}

\begin{figure}
\begin{center}
\includegraphics[scale=.6]{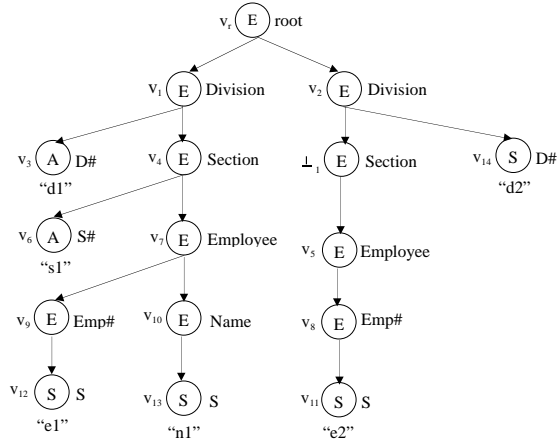}
\end{center}
\caption{The minimal extension of a XML tree. }
\label{F:Xcomplete}

\end{figure}

For instance, the minimal extension of the tree in Figure ~\ref{F:Xtree} is
shown
in Figure ~\ref{F:Xcomplete}.

\begin{definition}
A path instance $\bar{v}_1. \cdots. \bar{v}_n$ in $T$ is defined to be {\
\em complete}
if $\bar{v}_1. \cdots. \bar{v}_n \in T_P$. A tree $T$ is defined to be complete
w.r.t. a set of paths $P$ if $(T, P)$ is consistent and $T = T_P$. Also we
often do
not
need to distinguish between nulls and so the statement $v = \perp$ is
shorthand for
$\exists j(v = \perp_j)$ and $v \neq \perp$ is shorthand for $\not\exists j(v =
\perp_j)$.
\end{definition}

The next function returns all the final nodes of the path instances of a
path $p$.
\begin{definition}
Let $T_P$ be the minimal extension of $T$. The function $N(p)$, where $p$
is the
path $l_1.\cdots .
l_n$, is defined to be the set $\{\bar{v} | \bar{v}_1. \cdots. \bar{v}_n
\in Paths
(p)
\wedge \bar{v} = \bar{v}_n \}$.

\end{definition}

For example, in Figure ~\ref{F:Xcomplete},
$N(\texttt{root.Division.Section.Employee}) =\{v_7, v_5\}$ and

$N(\texttt{root.Division.Section }) = \{v_4, \perp_1\}$.

We now need to define a function that is related to ancestor.

\begin{definition}
Let $T_P$ be the minimal extension of $T$. The function $AAncestor(v, p)$
where $v \in V \cup {\bf N}$, $p$ is a path and $v
\in N(p)$,  is defined by $AAncestor(v, p) = \{v' | v' \in \{\bar{v}'_1,
\cdots,
\bar{v}'_n\} \wedge v = \bar{v}'_n \wedge \bar{v}'_1. \cdots. \bar{v}'_n \in
Paths(p)\}$.
\end{definition}

For example,  in Figure~\ref{F:Xcomplete}, $AAncestor (v_5,
\texttt{root.Division.Section.Employee}) = \{v_r, v_2, \perp_1, v_5\}$. The
next function returns all nodes that are the final nodes of path instances
of $p$ and
are descendants of $v$.

\begin{definition}
Let $T_P$ be the minimal extension of $T$.The function $Nodes(v, p)$, where $v
\in V \cup {\bf N}$ and $p$ is a path,
is the set defined by
$Nodes (v, p) = \{x | x \in N(p) \wedge  v \in AAncestor(x,p)\}$. Note that
$Nodes(v, p)$ may be
empty.
\end{definition}

For
example, in Figure  ~\ref{F:Xcomplete}, $Nodes(v_r,
\texttt{root.Division.Section.Employee}) = \{v_5, v_7\}$,

$Nodes(v_1, \texttt{root.Division.Section.Employee}) = \{ v_7\}$, $Nodes(v_7,
\texttt{root.Division}) = \phi$.

\begin{definition}
The partial ordering $>$ on the set of nodes $V$ in a XML tree $T$ is
defined by
$v_1 > v_2$ iff $v_2 \in Ancestor(v_1)$, where $v_1$ and $v_2$ are in $V$.
\end{definition}

In a similar fashion, we define a partial ordering on paths as follows.

\begin{definition}
The partial ordering $>$ on a set of paths $P$ is defined by $p_2 > p_1$ if
$p_1$ is a prefix of $p_2$, where $p_1$ and $p_2$ are paths in $P$.

\end{definition}
 For example, in Figure  ~\ref{F:Xcomplete}, \texttt{root.Division.D\#} $>$
\texttt{root.Division}. Also, \texttt{root.Division.D\#} and

\texttt{root.Division.Section} are incomparable.

Lastly we extend the definition of the $val$ function so that $val(\perp_j) =
\perp_j$.  Note that different unmarked nulls are not considered to be
equal and so
$val(\perp_i) \neq val(\perp_j)$ if $i \neq j$.

\section {Strong Functional Dependencies in XML}

 This leads us to the main definition of our paper.

\begin{definition}
\label{D:Xfd2}
Let $T$ be a XML tree and let $P$ be a set of paths such that $(T, P)$ is
consistent.
A XML functional dependency  (XFD) is a statement of the form:
$p_1, \cdots, p_k \rightarrow q$
where $p_1, \cdots, p_k $ and  $q$ are paths in $P$. $T$ {\em strongly
satisfies}
the XFD if $p_i = q$ for some $i, 1 \leq i \leq k$
or for any two distinct path instances
$\bar{v}_1.\cdots.\bar{v}_n$ and $\bar{v}'_1. \cdots. \bar{v}'_n$ in $Paths(q)$
in $T_P$,
(($\bar{v}_n = \perp \wedge
\bar{v}'_n = \perp ) \vee (\bar{v}_n \neq \perp \wedge \bar{v}'_n = \perp )
\vee
(\bar{v}_n = \perp \wedge
\bar{v}'_n
\neq \perp ) \vee (\bar{v}_n \neq \perp \wedge \bar{v}'_n \neq \perp \wedge
val(\bar{v}_n) \neq
val(\bar{v}'_n) \Rightarrow
\exists i, 1 \leq i \leq k$, such that $x_i \neq y_i$ if $Last(p_i) \in {\bf E}$ else
$\perp \notin Nodes(x_i, p_i) $ and $\perp
\notin Nodes(y_i, p_i) $ and $
val(Nodes(x_i,p_i)) \cap val(Nodes (y_i, p_i)) = \phi$,where $x_i =
\{v| v\in \{\bar{v}_1, \cdots, \bar{v}_n\}
\wedge v \in N(p_i \cap
q)\}$  and
$y_i = \{v | v \in \{v'_1, \cdots, \bar{v}'_n\}
\wedge v \in
N(p_i \cap q)\}$.
\end{definition}

We note that since the path $p_i \cap q$ is a prefix of $q$, there always
exists one
and only one node in $\{v'_1, \cdots, \bar{v}'_n\} $ that is also in $N(p_i
\cap q)$
and so $x_i$ is
always defined and unique.  Similarly for $y_i$.

We now outline the thinking behind the above definition firstly for the
simplest case
where the l.h.s. of the XFD contains a single path.  In the relational
model, if
we are given a relation $r$ and a FD $A \rightarrow B$, then to see if $A
\rightarrow
B$ is satisfied we have to check the $B$ values and their corresponding $A$
values.
In the relational model the correspondence between $B$ values and $A$ values is
obvious - the $A$ value corresponding to a $B$ value is the $A$ value in
the same
tuple as the $B$ value.  However, in XML there is no concept of a tuple so
it is not
immediately clear how to generalize the definition of an FD to XML. Our
solution
is based on the following observation. In a relation $r$ with tuple $t$,
the value
$t[A]$ can be seen as the 'closest' $A$ value to the $B$ value $t[B]$.  In
Definition
~\ref{D:Xfd2} we generalize this observation and given a path instance
$\bar{v}_1.
\cdots. \bar{v}_n$ in $Paths(q)$,  we first compute the 'closest' ancestor of $
\bar{v}_n$ that is also an ancestor of a node in $N(p)$ ($x_1$ in the above
definition) and then compute the 'closest p-nodes' to be the set of nodes which
terminate a path instance of $p$  and are descendants of $x_1$. We then
proceed in
a similar fashion for the other path $\bar{v}'_1. \cdots. \bar{v}'_n$ and
compute
the 'p-nodes' which are closest to $\bar{v}'_n$.  We note that in this
definition, as
opposed to the relational case, there will be in general more than one
'closest $p$ -
node' and so $Nodes(x_1, p)$ and $Nodes(y_1, p)$ will in general contain more
than one node.   Having computed the 'closest $p$-nodes' to $\bar{v}_n$ and
$\bar{v}'_n$, if $val(\bar{v}_n) \neq val(\bar{v}'_n)$ we then require,
generalizing on  the relational case, that the $val's$ of the sets of
corresponding
'closest $p$-nodes' be disjoint.

The rationale for the case where there is more than one path on the l.h.s.
is similar.
Given a XFD $p_1, \cdots, p_k \rightarrow q$ and two paths
$\bar{v}_1. \cdots. \bar{v}_n$ and $\bar{v}'_1. \cdots. \bar{v}'_n$ in
$Paths(q)$
which end in  nodes with different $val$, we firstly compute, for each
$p_i$, the
set of 'closest
$p_i$ nodes' to $\bar{v}_n$ in the same fashion as just outlined.
Then extending the relational approach to FD
satisfaction, we require that in order for $p_1, \cdots, p_k \rightarrow q$
to be
satisfied there is at least one $p_i$ for which the $val's$ of the set of
'closest $p_i$
nodes' to $\bar{v}_n$ is disjoint from the $val's$ of the set of 'closest
$p_i$ nodes'
to $\bar{v}'_n$. We now illustrate the definition by some
examples.

\begin{ex}

Consider the XML tree shown in Figure ~\ref{F:Xexamp} and the XFD

\texttt{root.Department.Lecturer.Subject.Subject\#} $\rightarrow$
\texttt{root.Department.Lecturer.Subject.SubjName.S}.  Then
$v_r.v_1.v_5.v_{13}.v_{17}.v_{22}$ and $v_r.v_2.v_9.v_{15}.v_{21}.v_{24}$
are two distinct path instances in

$Paths(\texttt{root.Department.Lecturer.Subject.SubjName.S})$ and $val(v_{22})
= \texttt{"n1"}$ and $val(v_{24}) = \texttt{"n2"}$.  So $
N(\texttt{root.Department.Lecturer.Subject.Subject\#} \cap$

$\texttt{root.Department.Lecturer.Subject.SubjName.S))} = \{v_{13}, v_{14},
v_{15}\}$ and so $x_1 = v_{13}$ and $y_1 = v_{15}$.  Thus $val(Nodes(x_1,
\texttt{root.Department.Lecturer.Subject.Subject\#})) = \{\texttt{"s1"} \}$ and

$val(Nodes(y_1, \texttt{root.Department.Lecturer.Subject.Subject\#})) =
\{\texttt{"s2"} \}$. Similarly for the paths
$v_r.v_1.v_6.v_{13}.v_{14}.v_{23}$ and $v_r.v_2.v_9.v_{15}.v_{21}.v_{24}$
and so the XFD is satisfied. We note that if we change $val$ of node
$v_{18}$ in
Figure ~\ref{F:Xexamp} to \texttt{"s1"} then the XFD is violated. \\

Consider next the XFD \texttt{root.Department.Head} $\rightarrow$
\texttt{root.Department}.  Then $v_r.v_1$ and $v_r.v_2$ are two distinct paths
instances in $Paths(\texttt{root.Department})$ and $val(v_1) = v_1$ and
$val(v_2)
= v_2$. Also

$N(\texttt{root.Department.Head } \cap \texttt{root.Department})  = \{v_{1},
v_{2} \}$ and so $x_1 = v_{1}$ and $y_1 = v_{2}$. Thus $val(Nodes(x_1,
\texttt{root.Department.Head})) = \{\texttt{"h1"} \}$ and $Val(Nodes(y_1,
\texttt{root.Department.Head})) = \{\texttt{"h2"} \}$ and so the XFD is
satisfied.
We note that if we change $val$ of node $v_8$ in Figure ~\ref{F:Xexamp} to
\texttt{"h1"} then the XFD is violated. \\

Consider next the XFD \texttt{root.Department.Lecturer.Lname},
\texttt{root.Department.Dname} $\rightarrow$

\texttt{root.Department.Lecturer.Subject.Subject\#}.  Then
$v_r.v_1.v_5.v_{13}.v_{16}$ and $v_r.v_2.v_9.v_{15}.\perp_1 $ are two
distinct path instances in
$Paths(\texttt{root.Department.Lecturer.Subject.Subject\#})$ and
$val(v_{16}) =
\texttt{"s1"}$ and the final node in $v_r.v_2.v_9.v_{15}.\perp_1 $  is null.

Then $N(\texttt{root.Department.Lecturer.Lname} \cap$ $
\texttt{root.Department.Lecturer.Subject.Subject\#})$
$= \{v_{5}, v_{6}, v_9 \}$
and so $x_1 = v_{5}$ and $y_1 = v_{9}$ and so $val(Nodes(x_1,
\texttt{root.Department.Lecturer.Lname})) = \texttt{"l1"}$ and
$val(Nodes(y_1, . \texttt{root.Department.Lecturer.Lname})) =
\texttt{"l1"}$. We
then compute

$N(\texttt{root.Department.Dname} \cap
\texttt{root.Department.Lecturer.Subject.Subject\#}) =\{v_1, v_2\}$ and so
$x_2 =
v_1$ and $y_2 = v_2$ and so $val(Nodes(x_2, . \texttt{root.Department.Dname}))
= \texttt{"d1"}$ and

$val(Nodes(y_2, . \texttt{root.Department.D name})) =
\texttt{"d2"}$.  Similarly, for the paths $v_r.v_1.v_6.v_{14}.v_{18}$, we
derive
that $x_2 = v_6$ and $y_2 = v_{9}$ and so

$val(Nodes(x_2, .\texttt{root.Department.Dname}))$ $\neq$
$val(Nodes(y_2, . \texttt{root.Department.D name})) = \texttt{"d2"}$ and so the
XFD is satisfied.
 Thus $x_1 \neq y_1$.  Similarly for $v_r.v_1.v_6$ and
$v_r.v_2.v_9.v_{15}.\perp_1 $ and so the XFD is satisfied.

\begin{figure}[here]
\begin{center}
\includegraphics[scale=.6]{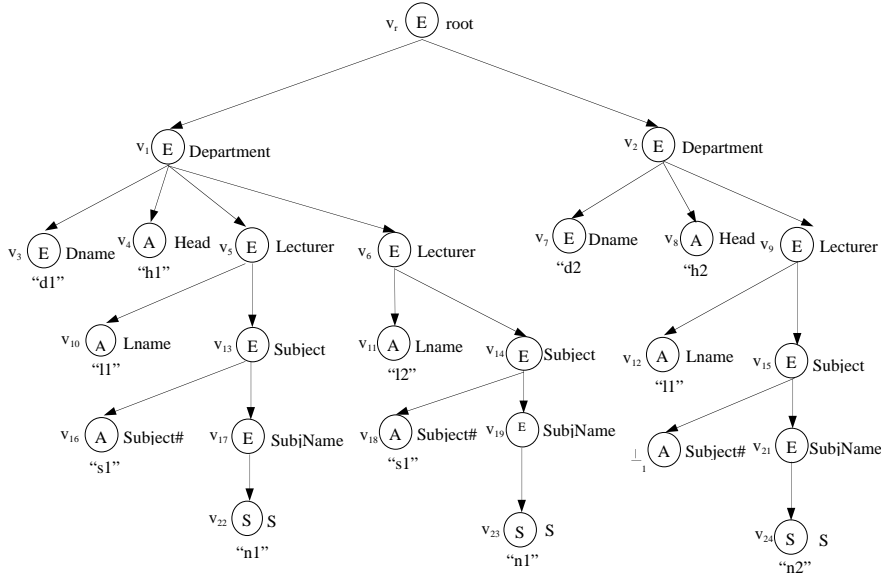}
\end{center}
\caption{A XML tree illustrating the definition of a XFD }
\label{F:Xexamp}

\end{figure}

\end{ex}

\section{Axiomatization for XFDs}
In this section we address the issues of completeness of the axiom system for
reasoning about implication of XFDs that was presented in \cite{Vin02a}.  The
axiom system is the following.

{\bf Axiom A1.} $p_1, \cdots, p_k \rightarrow p_i$ for any $p_i$, $1 \leq i
\leq
k$.

{\bf Axiom A2.} If $p_1, \cdots, p_k \rightarrow q$,  then $p, p_1, \cdots, p_k
\rightarrow q$ for any path $p$.

{\bf Axiom A3.} If $p_1, \cdots, p_k \rightarrow q$,  and  $q  \rightarrow s$
then  $p_1, \cdots, p_k \rightarrow s$.

{\bf Axiom A4.} If $p_1, \cdots, p_k \rightarrow q$ and $\forall i, 1 \leq
i \leq k,
p_i \cap q = root$, then $p \rightarrow q$  for any path
$p$.

{\bf Axiom A5.}  If $p \rightarrow q$ then $p' \rightarrow q$ for all paths
$p'$
such that $p \cap q$ is prefix of $p'$ and either $p'$ is a prefix of $p$
or $p'$
is a prefix of $q$.

{\bf Axiom A6.} If $ Last(p) \in {\bf E}$ and $q$ is a prefix of $p$ then $p
\rightarrow q$.

{\bf Axiom A7.} If $Last(q) \in {\bf A}$ then $Parnt(q) \rightarrow q$.

{\bf Axiom A8.} $p \rightarrow root$ for any path $p$.

\begin{theorem}
\label{T:sound}
Axioms A1 - A8 are sound for implication of arbitrary XFDs.
\end{theorem}
\
{\bf Proof.} For the sake of the completeness of this paper, the proof from
\cite{Vin02a} is reproduced in the Appendix.\\

We now illustrate these axioms by an example.

\begin{figure}[here]
\begin{center}
\includegraphics[scale=.6]{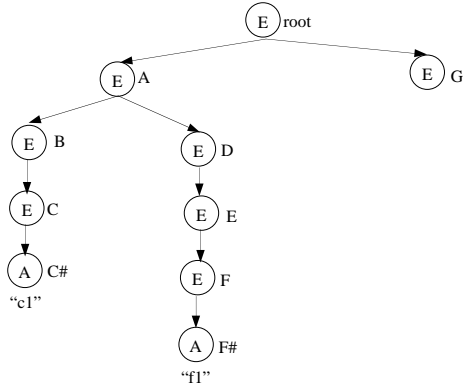}
\end{center}
\caption{XML tree illustrating axioms for XFDs. }
\label{F:Xaxiom}

\end{figure}

\begin{ex}

Consider the XML tree show in Figure ~\ref{F:Xaxiom} and the set $\Sigma$ of
XFDs \{\texttt{root.A.B.C.C\#} $\rightarrow$ \texttt{root.A.D.E},
\texttt{root.A.D.E} $\rightarrow$ \texttt{root.A.D.E.F.F\#}, \texttt{root.A}
$\rightarrow$ \texttt{root.G}\}.  It can be easily verified that the XML
tree in
Figure ~\ref{F:Xaxiom} satisfies $\Sigma$.  Then from $\Sigma$ and the axioms
we can deduce that the following XFDs are implied by $\Sigma$\footnote{We do
not show all the XFDs that can be derived from the axioms} : from A1 we can
derive \texttt{root.A} $\rightarrow$ \texttt{root.A}, from A2 and
\texttt{root.A}
$\rightarrow$ \texttt{root.G} we can derive that \texttt{root.A,}
\texttt{root.A.B.C} $\rightarrow$ \texttt{root.G}, from A3 and
\texttt{root.A.B.C.C\#} $\rightarrow$ \texttt{root.A.D.E} and
\texttt{root.A.D.E}
$\rightarrow$ \texttt{root.A.D.E.F.F\#} we can derive that
\texttt{root.A.B.C.C\#}
$\rightarrow$\texttt{root.A.D.E.F.F\#}, from A4 and \texttt{root.A}
$\rightarrow$ \texttt{root.G} we can derive that  \texttt{root.A.D.E}
$\rightarrow$ \texttt{root.G}, from A5 and \texttt{root.A.B.C.C\#}
$\rightarrow$
\texttt{root.A.D.E} we can derive that \texttt{root.A.B} $\rightarrow$
\texttt{root.A.D.E}  and that \texttt{root.A.D} $\rightarrow$
\texttt{root.A.D.E},
from A6 we can derive that \texttt{root.A.D.E} $\rightarrow$ \texttt{root.A},
from A7 we can derive that \texttt{root.A.D.E.F} $\rightarrow$
\texttt{root.A.D.E.F.F\#} and from A8 we derive that \texttt{root.A.D}
$\rightarrow$ \texttt{root}.
\end{ex}

This now leads to the first major result of the paper.

\begin{theorem}
\label{T:complete}
Axioms A1 - A8 are complete for unary XFDs
\end{theorem}

{\bf Proof.} See Appendix.

\section{Decidability Of Implication  for Unary XFDs}

In this section we derive the second main result of the paper by showing
that the
implication problem for unary XFDs is decidable. We do this by constructing an
algorithm for generating $P^+$, the set of all paths $q$ such that $ q \in
P^+$ if
and only if $p \rightarrow q \in \Sigma^+$ and then prove that the algorithm is
correct. We note also that the running time of the algorithm is linear in
the number
of XFDs in $\Sigma$. Firstly we present an algorithm which is analogous to the
classic chase procedure for relations \cite{Mai79}.

Before presenting the next algorithm, we define two  functions.

\begin{definition}
The function $Anc(p)$, where $p$ is a path, is the set defined by $Anc(p) =
\{q |
q$  is a {\em strict prefix} of $p\}$. The function $Att(p)$ is the set
defined by $Att(p) =
\{q | p = Parnt(q) \wedge Last(q) \in {\bf A} \}$.
\end{definition}

\pagebreak

\begin{Verbatim}[commandchars=\\\{\},codes={\catcode`$=3 \catcode`^=7}]
{\bf Algorithm 1}
INPUT: A set $\Sigma$ of unary  XFDs and a tree $T$ which is complete
w.r.t. the
set of paths in $\Sigma$.
OUTPUT: A XML tree $\bar{T}$ satisfying the set of XFDs and which is complete
w.r.t. the set of paths in $\Sigma$.
$\bar{T} = T$;
Repeat until no more changes can be made to $\bar{T}$
  For each $p \rightarrow q \in \Sigma$ do
    If $Last(q) \notin {\bf E}$ then
      If there exist $v1, v2, v3, v4 \in \bar{T}$ such that
        $v1, v2 \in N(p), v3, v4 \in N(q)$ and $val(v1) =  val(v2)$  and
        $val(v3) < val(v4)$ then
          $val(v4) := val(v3)$;
    If $Last(q) \in {\bf E}$ then
      If there exist $v1, v2, v3, v4$ in $\bar{T}$ such that
        $v1, v2 \in N(p), v3, v4 \in N(q)$ and $val(v1) =  val(v2)$  then
          attach all descendants of $v4$ to $v3$;
          DeleteSameAtts(v3);
          vl := Parent(v3); vr := Parent(v4);
          repeat until vl = vr
            attach all descendants of vr to vl except for v4;
            deleteSameAtts(vl);
            delete(v4);
            v4 := vr;
            vl:= Parent(vl); vr := Parent(vr);
          endrepeat;
  endfor
endrepeat

procedure DeleteSameAtts (node v);
   For any pair of nodes $v5$ and $v6$ such that $v5$ and $v6$  are children
   of $v$  and $lab(v5) = lab(v6)$ and $lab(v5) \in {\bf A}$ and
   $val(v5) \leq val(v6)$ then  delete $v6$;
return;

\end{Verbatim}

We now illustrate Algorithm 1 by an example.

\begin{figure}[here]
\begin{center}
\includegraphics[scale=.6]{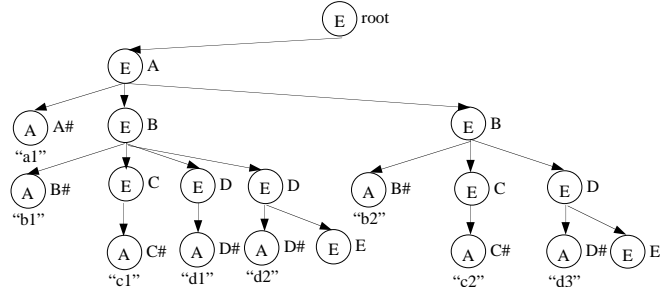}
\end{center}
\caption{Initial XML tree }
\label{F:Tchase1}
\end{figure}

\begin{figure}[here]
\begin{center}
\includegraphics[scale=.6]{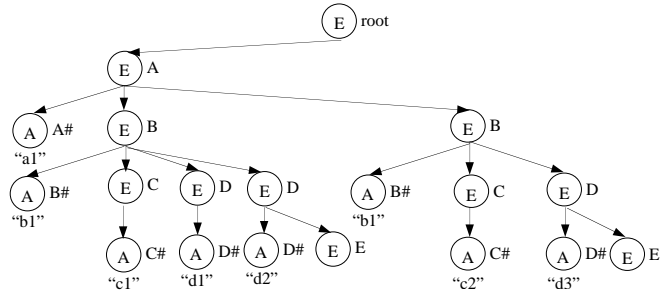}
\end{center}
\caption{ XML tree after applying  $\texttt{root.A.A\#}$ $ \rightarrow$
\texttt{root.A.B.B\#}}
\label{F:Tchase2}
\end{figure}

\begin{figure}[here]
\begin{center}
\includegraphics[scale=.6]{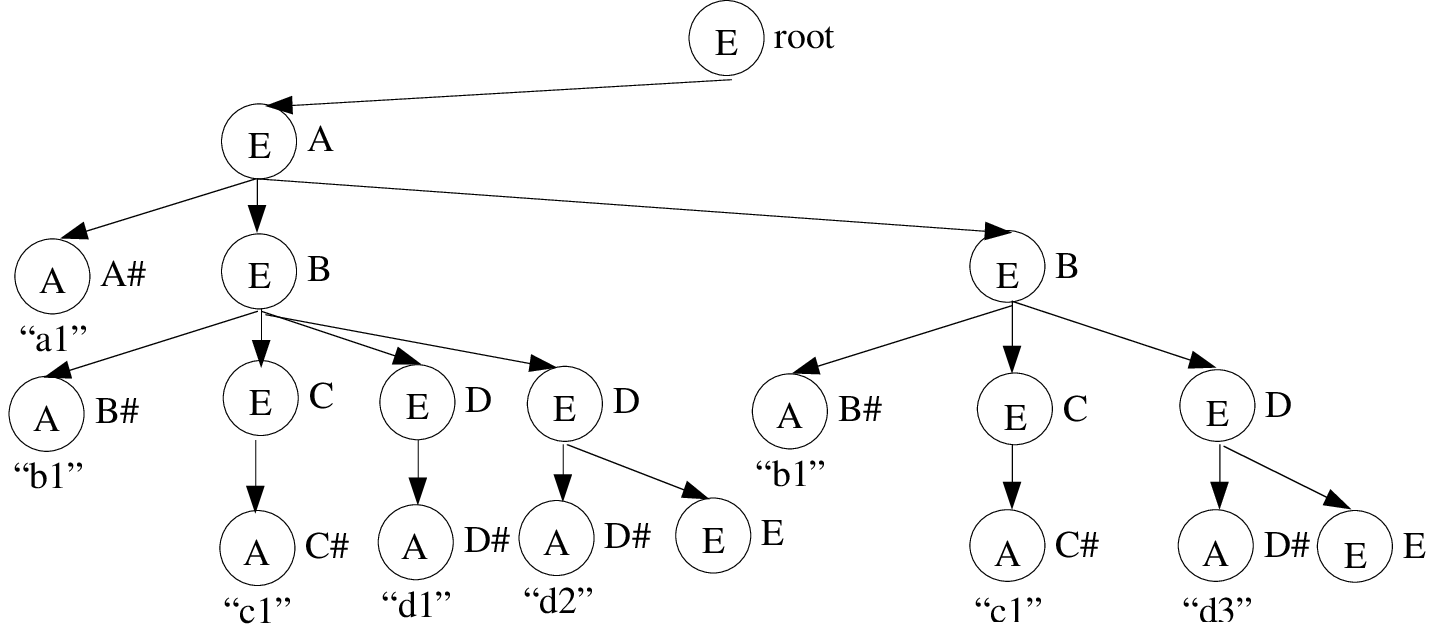}
\end{center}
\caption{XML tree after applying \texttt{root.A.B.B\#} $\rightarrow$
\texttt{root.A.B.C.C\#} }
\label{F:Tchase3}
\end{figure}

\begin{figure}[here]
\begin{center}
\includegraphics[scale=.6]{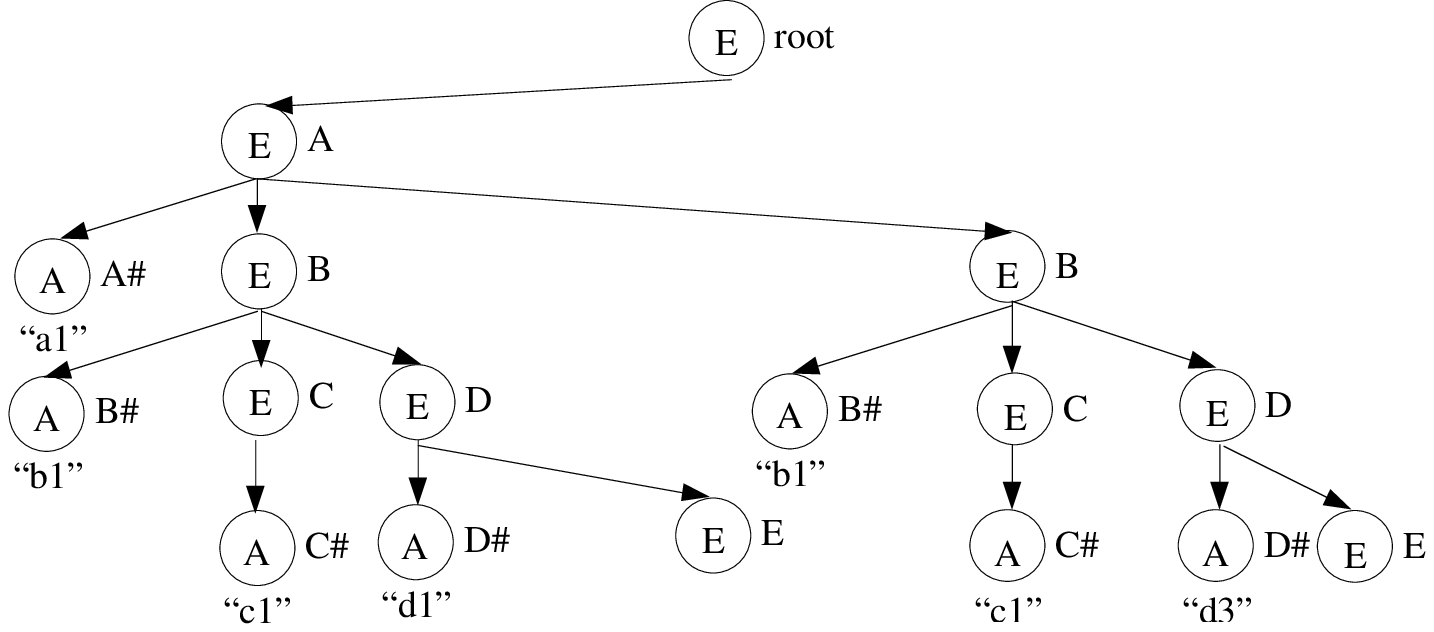}
\end{center}
\caption{XML tree after applying \texttt{root.A.B} $\rightarrow$
\texttt{root.A.B.D}}
\label{F:Tchase4}
\end{figure}

\begin{ex}
Let $\Sigma$ be the set of XFDs $\{\texttt{root.A.A\#}$ $ \rightarrow$
\texttt{root.A.B.B\#}, \texttt{root.A.B.B\#} $\rightarrow$
\texttt{root.A.B.C.C\#},
\texttt{root.A.B} $\rightarrow$ \texttt{root.A.B.D}\} and let the initial
tree $T$ be
as shown in Figure ~\ref{F:Tchase1}. Then if we apply the XFD
$\texttt{root.A.A\#}$ $ \rightarrow$ \texttt{root.A.B.B\#} the resulting
tree is
shown in Figure ~\ref{F:Tchase2}.  If we then apply \texttt{root.A.B.B\#}
$\rightarrow$ \texttt{root.A.B.C.C\#} the resulting tree is shown in Figure
~\ref{F:Tchase3}.  Finally, if we apply \texttt{root.A.B} $\rightarrow$
\texttt{root.A.B.D} then the tree is shown in Figure ~\ref{F:Tchase4}.
\end{ex}

\begin{lemma}
Algorithm1 always terminates.
\end{lemma}

{\bf Proof.} The function $ Count(p)$, where $p$ is a path and $Last(p) \in
{\bf
E}$, is defined to be $|N(p)|$, were $| |$ denotes cardinality.  The
function $Sum(p)$, where $p \notin {\bf E}$ is
defined as follows.   For any text string value \texttt{t} define the function
$int(\texttt{t})$ to be the integer value obtained by considering
\texttt{t} to be an
integer to base 256 and then define $Sum(p)$  to be $\sum_{v \in N(p)}
int(val(v))$. At each iteration of the repeat loop either $Count(p)$ or
$Sum(p)$
strictly decreases for  at least one path $p$.  Hence since both $Count(p)$ and
$Sum(p)$ are both bounded below by 0 Algorithm 1 must terminate.
\hfill$\Box$\\

Firstly, let us denote by $P_\Sigma$ the set of paths that appear on the
l.h.s. or r.h.s. of any XFD in a set of unary XFDs $\Sigma$.

\begin{lemma}
\label{L:complete}
The tree $\bar{T}$ produced by Algorithm 1 is complete w.r.t. $P_\Sigma$.
\end{lemma}
{\bf Proof.} The proof is by induction. Initially the result is true
because of the
restriction placed on the input tree $T$ by Algorithm 1. Assume then the
result is
true after iteration $k-1$.  Then during iteration $k$, the only path
instances which
can possibly be changed are those in $Paths(q)$ or $Paths(Anc(q))$ for some
$q \in
{\bf E}$. However, if we merge two nodes in $N(q)$  then we also merge their
ancestor nodes  and so after iteration $k$, $Paths(q)$ and $Paths(Anc(q))$ will
again contain only complete paths and so the result is established.
\hfill$\Box$\\

\begin{lemma}
\label{L:term}
The tree generated by Algorithm 1 satisfies $\Sigma$.
\end{lemma}

{\bf Proof.}
From the definition of the algorithm, the algorithm terminates only when
there is
no XFD that is violated.
\hfill$\Box$\\

Next, we introduce an algorithm for calculating the closure of a set of XFDs.

\pagebreak
\begin{Verbatim}[commandchars=\\\{\},codes={\catcode`$=3 \catcode`^=7}]
{\bf Algorithm 2}
INPUT: A set $\Sigma$ of unary XFDs and a path $p$.
OUTPUT:  $P^+$ the set of paths such that $q \in P^+$ iff
                 $p \rightarrow q $ is implied by $\Sigma$.
  1:$P^+ = \{p, root\}$;
  2:If $Last(p) \in {\bf E}$ then $P^+ = P^+ \cup Anc(p)$;
  3: for each $q \in P^+$ do $P^+ := P^+ \cup Att(q)$;
  $Unused = \Sigma$;
  repeat until no more changes to $P^+$
    Choose arbitrarily $r \rightarrow s$ from $\Sigma$;
    4:(If $\exists p1 \in P^+$ such that $r \cap s$ is prefix of $p1$ and
    either $p1$ is a prefix of $r$ or $p1$ is a prefix of $s$)
    5:$\vee (r \in P^+)$
    6:$\vee (r \cap s = root)$ then
      $P^+ = P^+ \cup \{s\};$
      $Unused = Unused - \{r \rightarrow s\}$;
      7:if $Last(s) \in {\bf E}$ then $P^+ = P^+ \cup Anc(s) \cup Att(s)$;
  endrepeat
\end{Verbatim}

We note that it is easily seen that since each XFD in $\Sigma$ is used only
once, the
running time of Algorithm 2 is linear in the number of XFDs in $\Sigma$.  We
now proceed to prove that Algorithm 2 is correct.  Two cases are considered
separately.

\underline{Case 1: $Last(p) \notin {\bf E}$}

First construct a tree $T_0$ with the following properties.  $T_0$ is complete
w.r.t. $P_\Sigma$ and for every path  appearing in $P_\Sigma$, except for the
root, there are exactly two path instances of the path in $T_0$. Also, the path
instances for $p$ have the property that the $val$ of the final nodes in
the path
instances are the same whereas the $val$ of the end nodes for the path
instances of
any other path in $P_\Sigma$ are distinct.  Such a tree can always be
constructed.
We now illustrate the construction by an example.

\begin{ex}
Let $\Sigma = \{\texttt{root.A.B.B\#} \rightarrow \texttt{root.A.A\#},
\texttt{root.A.C.C\#} \rightarrow \texttt{root.A.B}, \texttt{root.A.A\#}
\rightarrow \texttt{root.D.D\#}\}$ and let $p$ be the path
\texttt{root.A.B.B\#}.
Then the tree $T_0$ is shown in Figure ~\ref{F:T0}.

\begin{figure}[here]
\begin{center}
\includegraphics[scale=.6]{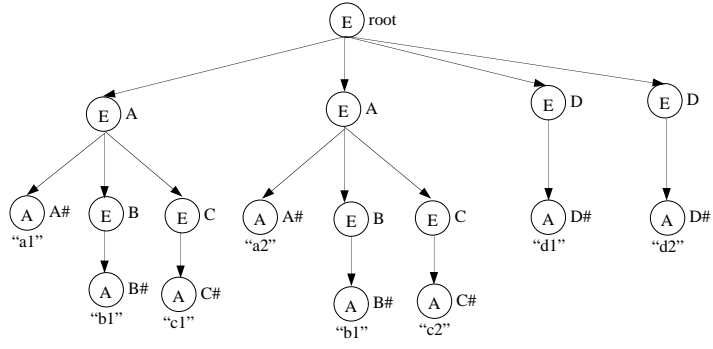}
\end{center}
\caption{A XML tree }
\label{F:T0}

\end{figure}

\end{ex}
The next step is using as input the set of XFDs returned in $P^+$ by
Algorithm 2
and the tree $T_0$, generate the tree $\bar{T}_0$ using Algorithm 1.  We note
that it follows from Lemma ~\ref{L:term} that $\bar{T}_0$ satisfies  $P^+$. We
now prove some preliminary lemma before establishing the main result.

\begin{lemma}
\label{L:8}
Let $\bar{v}_1. \cdots. \bar{v}_n$ and $\bar{v}'_1. \cdots. \bar{v}'_n$ be  two
distinct path instances in  $Paths(q)$ in $T_0$ for any path $q$ in
$P_\Sigma$.
Then the only common node to both path instances is $root$.
\end{lemma}

{\bf Proof.}  Suppose to the contrary that node $ \bar{v}_j$ is common to both
path instances.  Then $ \bar{v}_j \in N(s)$ for some path $s$ that is a
prefix of
$q$.  So because of the definition of $T_0$ there must exist two path
instances in
$Paths(s)$ and so there exists another node $ \bar{v}^1_j$ in $N(s)$ that
is distinct
from $ \bar{v}_j$.  There are then two possibilities.  The first is that
there exists
another path instance $\bar{v}''_1. \cdots. \bar{v}''_n$ in $Paths(q)$ that
contains
$ \bar{v}^1_j$.  If this is the case then since $ \bar{v}^1_j$ and $
\bar{v}_j$ are
distinct the paths
$\bar{v}_1. \cdots. \bar{v}_n$,  $\bar{v}'_1. \cdots. \bar{v}'_n$ and
$\bar{v}''_1. \cdots. \bar{v}''_n$ are distinct which contradicts the fact
that $T_0$
has only two path instances for any path.  The other possibility is that
there is no
path instance in $Paths(q)$ in $T_0$ that contains $ \bar{v}^1_j$ but this
contradicts the fact that $T_0$ is complete w.r.t. $P_\Sigma$.  So either
possibility
leads to a contradiction so we conclude that the only common node to
$\bar{v}_1.
\cdots. \bar{v}_n$ and $\bar{v}'_1. \cdots. \bar{v}'_n$ is $root$.
\hfill$\Box$\\

\begin{lemma}
\label{L:1}
Let $q$ be any path in $P_\Sigma$. Then if there exist two distinct path
instances
$\bar{v}_1. \cdots. \bar{v}_n$ and $\bar{v}'_1. \cdots. \bar{v}'_n$ in
$Paths(q)$
in $\bar{T}_0$ such that $\bar{v}_1. \cdots. \bar{v}_n$ and $\bar{v}'_1.
\cdots.
\bar{v}'_n$  have a common node that is not the $root$ then there exists
$s'$ such
that $s' \in Anc(q)$ and $s' \in P^+$.

\end{lemma}

{\bf Proof.} We prove the result by induction on the number of steps in
constructing $\bar{T}_0$. Initially the result is true for $T_0$ by Lemma
~\ref{L:8}.  Assume inductively then that it is true after iteration $k-1$.
The only
way that we can have that $\bar{v}_1. \cdots. $ and $\bar{v}'_1. \cdots. \bar{v}'_n$
in $Paths(q)$  have a common non root node after iteration $k$ is if we
merge two
ancestor nodes  of $\bar{v}_n$ and $\bar{v}'_n$. For this to happen we have to
have $s$ such that $s \in Anc(q)$ and $s \in P^+$.
\hfill$\Box$\\

\begin{lemma}
\label{L:4}
If a XFD $r \rightarrow s$ is violated in $\bar{T}_0$ then $p \rightarrow s$ is
violated in $\bar{T}_0$.
\end{lemma}

{\bf Proof.} If $r \rightarrow s$ is violated then  there exist distinct
path instances
$\bar{v}_1. \cdots. \bar{v}_n$ and $\bar{v}'_1. \cdots. \bar{v}'_n$ in
$Paths(s)$
such that $ val(\bar{v}_n) \neq val(\bar{v}'_n)$.  However by the
construction of
$\bar{T}_0$, $N(p)$ contains only two nodes, say $v_1$ and $v_2$, such that
$val(v_1) = val(v_2)$. Let us then compute $x_1 =
\{v| v \in \{\bar{v}_1, \cdots, \bar{v}_n\} \wedge v \in N(p \cap s)\}$  and
$y_1 =  \{v | v \in \{v'_1, \cdots, \bar{v}'_n\} \wedge v \in
N(p \cap s) \}$.   If $x_1 = y_1$ then $Nodes(x_1, p) = Nodes(y_1, p)$ and so
$Nodes(x_1, p) \cap Nodes(y_1, p) \neq \phi$ and so $p \rightarrow s$ is
violated.
If $x_1 \neq y_1$ then we must have that $val(Nodes(x_1, p)) \cap
val(Nodes(y_1,
p)) \neq \phi$ because$v_1 \in Nodes(x_1, p)$ and  $v_2 \in Nodes(y_1, p)$ and
$val(v_1) = val(v_2)$ and so $p \rightarrow s$ is again violated.
\hfill$\Box$\\

\begin{lemma}
\label{L:2}
If there is a path $q$ in $P_\Sigma$ such that $Last(q) \notin {\bf E}$
then $q \in
P^+$ iff there exist two distinct nodes $v_1$ and $v_2$ in $N(q)$ in
$\bar{T}_0$
such that $val(v_1) = val(v_2)$.
\end{lemma}

{\bf Proof.}

{\it If:} We prove the result again by induction on the number of steps in
constructing $\bar{T}_0$.  Initially the result is true since $p \in P^+$
and the
$val$ of the two nodes in $N(p)$ is the same. Suppose then that  there
exist two
nodes in in $N(q)$ such that $val(v_1) \neq val(v_2)$ before step $k$ and
$val(v_1) = val(v_2)$ after step $k$.  By the definition of Algorithm 1 the
only
way for this to happen is if $q \in P^+$.

{\it Only If:}  We shall show the contrapositive that if there exist two
nodes $v_1$
and $v_2$ in $N(q)$ such that $val(v_1) \neq val(v_2)$ then $q \notin P^+$. If
there exist two nodes $v_1$ and $v_2$ in $N(q)$ such that $val(v_1) \neq
val(v_2)$ then using the same reasoning as in Lemma ~\ref{L:4} it follows that
$\bar{T}_0$ violates $p \rightarrow q$ and so by Lemma ~\ref{L:term} we must
have that $q \notin P^+$.
\hfill$\Box$\\

\begin{lemma}
\label{L:3}
If there is a path $q$ in $P_\Sigma$ such that $Last(q) \in {\bf E}$ then
$q \in
P^+$ iff $|N(q)| = 1$ in $\bar{T}_0$.
\end{lemma}

{\bf Proof.}

{\it If:}  We prove the result by induction on the number of steps in
generating
$\bar{T}_0$.  Initially the result is true since $p$ is the only node in
$P^+$ and
$Last(p) \notin {\bf E}$.  Suppose then it is true after iteration $k$.
Then by
definition of the algorithm, the only  case where we can have a  path $q$
such that
$|N(q)| \neq 1$ after step $k$ but $|N(q)| = 1$ after step $k+1$ is if $q
\in P^+$.

{\it Only If:}  Suppose $|N(q)| \neq 1$. By the construction of $\bar{T}_0$
it can
easily be seen that $N(p)$ contains only two nodes and $val$ of the nodes
are equal.
Thus  using the same arguments as used in Lemma ~\ref{L:4} that $p \rightarrow
q$ is violated in $\bar{T}_0$ which is a contradiction since $q \in P^+$ and by
Lemma ~\ref{L:term} $\bar{T}_0$ satisfies $p \rightarrow q$. Hence we conclude
that $|N(q)| = 1$.
\hfill$\Box$\\

\begin{theorem}
\label{T:1}

Algorithm 2 correctly computes $P^+$ when $Last(p) \notin {\bf E}$.
\end{theorem}

{\bf Proof.}   We firstly show that if $q \in P^+$  then $p \rightarrow q$
is in
$\Sigma^+$.  We show this by induction on the number of iterations in computing
$P^+$.  At line 1 $P^+$ contains $p$ and $root$ and  $p \in P^+$ by axiom
A1 and
$root \in P^+$ by axiom A8.  At line 2  the result follows by axiom A6 and
at line
3 by axiom A7.  Hence the inductive hypothesis is true at the commencement
of the
loop.  Let $P^+_j$ denote the computation of $P^+$ after iteration $j$. Assume
then that the hypothesis is true after iteration $j-1$. If the $q$ is added
to $P^+_j$
because of line 4 then $p \rightarrow q$ is in $P^+$ because of axiom A5
and the
induction hypothesis.  If $q$ is added at line 5 then $r \in P^+$ by the
induction
hypothesis  and axiom A3. If $q$ is added because line 6 then $q \in P^+$
by axiom
A4.  If $q$ is added as a result of line 7 then $q \in P^+_j$ because of
axioms A6
and A7.

Next we show that if $p \rightarrow q \in \Sigma^+$ then $q \in P^+$. We
firstly
claim that $\bar{T}_0$ satisfies $\Sigma$ (note that this does not follows from
Lemma ~\ref{L:term} since we are using $P^+$ as input to Algorithm 1 rather
than $\Sigma$). Let $r \rightarrow s$ be any XFD in $\Sigma$.  Suppose firstly
that $r = root$. If $r \rightarrow s$ is violated in $\bar{T}_0$ then by Lemma
~\ref{L:4} $p \rightarrow s$ must be violated. However since $root
\rightarrow s
\in \Sigma$ then we must have that $s \in P^+$ or else $s$ could be added
at line 6
contradicting the definition of $P^+$. So by Lemma ~\ref{L:term}  $p
\rightarrow
s$ is satisfied in $\bar{T}_0$ hence we conclude that $r \rightarrow s$ is
satisfied
in $\bar{T}_0$ or else by Lemma ~\ref{L:4} $p \rightarrow s$ is violated
which is
a contradiction.
 Suppose then that $r \rightarrow s$ is violated in $\bar{T}_0$ and $r \neq
root$.
The first way for this to happen is if  there exist two path instances
$\bar{v}_1.
\cdots. \bar{v}_n$ and $\bar{v}'_1. \cdots. \bar{v}'_n$ in $Paths(s)$ such that
$x_1 \neq y_1$, where $x_1 =
\{v| v \in \{\bar{v}_1, \cdots, \bar{v}_n\} \wedge v \in N(r \cap
s) \}$  and
$y_1 = \{v | v \in \{v'_1, \cdots, \bar{v}'_n\} \wedge v \in
N(r \cap s) \}$, and there exists $v_1 \in Nodes(x_1, r)$ and $v_2 \in
Nodes(y_1,
r)$ such that $val(v_1) = val(v_2)$.   For this to happen it follows from Lemma
~\ref{L:2} that $r \in P^+$.  We must also have that $s \in P^+$ or else
$s$ could
be added to $P^+$ by line 5 thus contradicting the definition of $P^+$.
However,
if $s \in P^+$ then by Lemma ~\ref{L:term} $p \rightarrow s$ is satisfied in
$\bar{T}_0$ which contradicts the assumption  that $r \rightarrow s$ is
violated in
$\bar{T}_0$ by Lemma ~\ref{L:4}.  We conclude that in this case $r \rightarrow
s$ is satisfied.   The second way that $r \rightarrow s$ could be violated in
$\bar{T}_0$  is if there exist two path instances $\bar{v}_1. \cdots.
\bar{v}_n$
and $\bar{v}'_1. \cdots. \bar{v}'_n$ in $Paths(s)$ such that $x_1 = y_1$.
If $x_1
= root$ then $r \cap s = root$ and so $s \in P^+$ or else it could be added
at line 6
contradicting the definition of $P^+$. If $r \rightarrow s$ is violated,
then by
Lemma  ~\ref{L:4} $ p \rightarrow s$ is violated which contradicts Lemma
~\ref{L:term} and so we conclude that $r \rightarrow s$ is satisfied.
Assume then
that $x_1 \neq root$ and so
by Lemma ~\ref{L:1} and since $x_1 = y_1$ there exists $s'$ such that $s'
\in Anc(s)$ and $s' \in P^+$. It
follows that $r \cap s$ is a prefix of $s'$ and so $r \cap s \in Anc(s)$
and thus $r
\cap s \in P^+$ or else  it could be added at line 7 of Algorithm 2 which
contradicts
the definition of $P^+$. Next, since $r \cap s \in P^+$ it follows that
$s$ must be
in $P^+$ or else it could be added at line 4 since $r \cap s$ is a prefix
of $r$ which
contradicts the definition of $P^+$. However, by Lemma ~\ref{L:term}
$\bar{T}_0$ satisfies $p \rightarrow s$ and if $r \rightarrow s$ is violated in
$\bar{T}_0$ then $p \rightarrow s$ is violated in $\bar{T}_0$ by Lemma
~\ref{L:4} which is a contradiction and so $r \rightarrow s$ must be
satisfied in
$\bar{T}_0$.

To complete the proof suppose that  $p \rightarrow q \in \Sigma^+$. Since
$\bar{T}_0$ satisfies $\Sigma$  then $\bar{T}_0$ also satisfies $p
\rightarrow q$.
Suppose firstly that $Last(q) \in {\bf E}$.  If $|N(q)| \neq 1$ then using
a similar
argument to Lemma ~\ref{L:4}  this would imply that $p \rightarrow q$ is
violated
in $\bar{T}_0$ which is a contradiction and so $|N(q)| = 1$. Then by Lemma
~\ref{L:3} $q \in P^+$. Suppose instead that that $Last(q) \notin {\bf E}$. By
definition of Algorithm 1, there exists two nodes in $N(q)$, say $v_1$ and
$v_2$,
since $Last(q) \notin {\bf E}$ and non element nodes are not removed in the
Algorithm. If $val(v_1) \neq val(v_2)$ then using similar arguments to
those used
in Lemma  ~\ref{L:4} it follows that $\bar{T}_0$ violates $p \rightarrow q$
which is a contradiction. Thus we must have that $val(v_1) = val(v_2)$ and
so by
Lemma ~\ref{L:2} $q \in P^+$.  This completes the proof.
\hfill$\Box$\\

\underline{Case 2: $Last(p) \in {\bf E}$}

First construct a tree $T_1$ with the following properties.  $T_1$ is complete
w.r.t. $P_\Sigma$.  For $p$ and every path $q$ such that $q$ is a prefix of
$p$ or
$q$ is an attribute node and its parent is a prefix of $p$, $T_1$ contains
exactly
one path instance.  For every other path in $P_\Sigma$, $T_1$ contains
exactly two
path instances. Also, the $val$ of any node in $T_1$ is distinct.  Such a
tree always
exists.  The construction procedure is illustrated by the following example.

\begin{ex}
Let $\Sigma = \{\texttt{root.A.B.B\#} \rightarrow \texttt{root.A.B.C.C\#},
\texttt{root.A.B.C.C\#} \rightarrow \texttt{root.A.A\#}, \texttt{root.A.D.D\#}
\rightarrow \texttt{root.E.E\#}\}$ and let $p$ be the path
\texttt{root.A.B}.  Then
the tree $T_1$ is shown in Figure ~\ref{F:T1}.

\begin{figure}[here]
\begin{center}
\includegraphics[scale=.6]{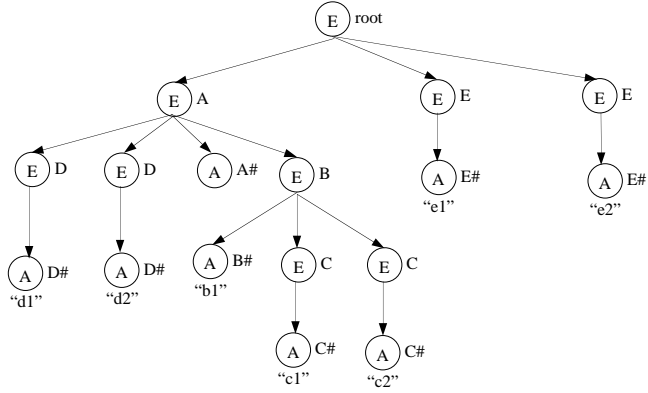}
\end{center}
\caption{A XML tree }
\label{F:T1}

\end{figure}
\end{ex}

The next step is using as input the set of XFDs returned in $P^+$ by
Algorithm 2
and the tree $T_1$ generate the tree $\bar{T}_1$ using Algorithm 1.  We
note that
it follows from Lemma ~\ref{L:term} that $\bar{T}_1$ satisfies  $P^+$. We now
prove some preliminary lemmas before establishing the main result.

\begin{lemma}
\label{L:b8}
Let $\bar{v}_1. \cdots. \bar{v}_n$ and $\bar{v}'_1. \cdots. \bar{v}'_n$ be two
distinct path instances in  $Paths(q)$ in $T_1$ for any path $q$ appearing in
$P_\Sigma$.  Then $\bar{v}_1. \cdots. \bar{v}_n$ and $\bar{v}'_1. \cdots.
\bar{v}'_n$ have a common node that is not the root only if $q \cap p \neq
root$.
\end{lemma}

{\bf Proof.}  Suppose to the contrary that $\bar{v}_1. \cdots. \bar{v}_n$ and
$\bar{v}'_1. \cdots. \bar{v}'_n$ have a common node $\bar{v}_j$ that is not the
root and $q \cap p = root$.  Then $ \bar{v}_j \in N(s)$ for some path $s$
such that
$s$ is a prefix of $q$.  So because of the definition of $T_1$ there must
exist two
path instances in $Paths(s)$ and so there exists another node $ \bar{v}^1_j$ in
$N(s)$ that is distinct from $ \bar{v}_j$. Then since $q \cap p = root$, it
follows
that $s$ is not a prefix of $p$.  Hence there must be two path instances of
$s$ in
$T_1$ and using the same reasoning as in Lemma ~\ref{L:8} shows that this leads
to a contradiction.
\hfill$\Box$\\

\begin{lemma}
\label{L:b1}
Let $q$ be any path in $P_\Sigma$. Then if there exist two distinct path
instances
$\bar{v}_1. \cdots. \bar{v}_n$ and $\bar{v}'_1. \cdots. \bar{v}'_n$ in
$Paths(q)$
in $\bar{T}_1$ such that $\bar{v}_1. \cdots. \bar{v}_n$ and $\bar{v}'_1.
\cdots.
\bar{v}'_n$  have a common node that is not the $root$ then either (there
exists
$s'$ such that $s' \in Anc(q)$ and $s' \in P^+$) or $q \cap p \neq root$.

\end{lemma}

{\bf Proof.} We prove the result by induction on the number of steps in
constructing $\bar{T}_1$. Initially the result is true for $T_0$ by Lemma
~\ref{L:b8}.  Assume inductively then that it is true after iteration
$k-1$. The only
way that we can have that $\bar{v}_1. \cdots. $ and $\bar{v}'_1. \cdots.
\bar{v}'_n$ in $Paths(q)$  have a common non root node after iteration $k$
is if
we merge two ancestor nodes  of $\bar{v}_n$ and $\bar{v}'_n$. For this to
happen
by definition of Algorithm 1 we have  that $s$ such that $s \in Anc(q)$ and
$s \in
P^+$.
\hfill$\Box$\\

\begin{lemma}
\label{L:b4}
If a XFD $r \rightarrow s$ is violated in $\bar{T}_1$ then $p \rightarrow s$ is
violated in $\bar{T}_1$.
\end{lemma}

{\bf Proof.} If $r \rightarrow s$ is violated then  there exist distinct paths
$\bar{v}_1. \cdots. \bar{v}_n$ and $\bar{v}'_1. \cdots. \bar{v}'_n$ in
$Paths(s)$
such that $ val(\bar{v}_n) \neq val(\bar{v}'_n)$.  However by the
construction of
$\bar{T}_1$, $N(p)$ contains only one node and so if we compute $x_1 =
\{v| v \in \{\bar{v}_1, \cdots, \bar{v}_n\} \wedge v  \in N(p \cap
s)\}$  and
$y_1 = \{v | v \in \{v'_1, \cdots, \bar{v}'_n\} \wedge v \in
N(p \cap s)\}$. Then $x_1 = y_1$ since $T$ is a tree and so by definition  of a
XFD $p \rightarrow s$ is violated.
\hfill$\Box$\\

\begin{lemma}
\label{L:b2}
If there is a path $q$ in $P_\Sigma$ such that $Last(q) \notin {\bf E}$
then $q \in
P^+$ iff there exist two distinct nodes $v_1$ and $v_2$ in $N(q)$ in
$\bar{T}_1$
such that $val(v_1) = val(v_2)$.
\end{lemma}

{\bf Proof.}

{\it If:} We prove the result again by induction on the number of steps in
constructing $\bar{T}_1$.  Initially the result is true since there is no
path $q$ and
two nodes $v_1$ and $v_2$ in $N(q)$ in $\bar{T}_1$ such that $val(v_1) =
val(v_2)$.  Assume then it is true after iteration $k - 1$.   Suppose then
that  there
exist two nodes in in $N(q)$ such that $val(v_1) \neq val(v_2)$ before step
$k$ and
$val(v_1) = val(v_2)$ after step $k$.  By the definition of Algorithm 1 the
only
way for this to happen is if $q \in P^+$.

{\it Only If:}  We shall show the contrapositive that if there exist two
nodes $v_1$
and $v_2$ in $N(q)$ such that $val(v_1) \neq val(v_2)$ then $q \notin P^+$. If
there exist two nodes $v_1$ and $v_2$ in $N(q)$ such that $val(v_1) \neq
val(v_2)$ then using the same reasoning as in Lemma ~\ref{L:b4} it follows that
$\bar{T}_1$ violates $p \rightarrow q$ and so by Lemma ~\ref{L:term} we must
have that $q \notin P^+$.
\hfill$\Box$\\

\begin{lemma}
\label{L:b3}
If there is a path $q$ in $P_\Sigma$ such that $Last(q) \in {\bf E}$ then
$q \in
P^+$ iff $|N(q)| = 1$ in $\bar{T}_1$.
\end{lemma}

{\bf Proof.}

{\it If:}  We prove the result by induction on the number of steps in
generating
$\bar{T}_1$.  Initially, by definition of $T_1$, if $|N(q)| = 1$ then
either $q = p$
or $q$ is a prefix of $p$  or $Last(q) \in {\bf A}$ and the $Parnt(q)$ is a
prefix of
$p$.  If $q = p$ then $q \in P^+$ by line 1. If $q$ is a prefix of $p$
then $q \in
Anc(p)$ and so $q \in P^+$ by line 2. If $Last(q) \in {\bf A}$ and the
$Parnt(q)$ is
a prefix of $p$ then $q \in P^+$ by line 3. Hence at the start of the
repeat loop the
result is true. Suppose then it is true after iteration $k$.   Then by
definition of the
Algorithm 1, the only  case where we can have a  path $q$ such that $|N(q)|
\neq 1$
after step $k$ but $|N(q)| = 1$ after step $k+1$ is if $q \in P^+$.

{\it Only If:}  Suppose to the contrary that $|N(q)| \neq 1$. By the
construction of
$\bar{T}_1$ it can easily be seen that $N(p)$ contains only one node.
Hence if we
define $x_1 =
\{v| v \in \{\bar{v}_1, \cdots, \bar{v}_n\} \wedge v \in N(p \cap
q))\}$  and
$y_1 =  \{v | v \in \{v'_1, \cdots, \bar{v}'_n\} \wedge v  \in
N(p \cap q) \}$ then $x_1 = y_1$ since $|N(p)| = 1$ and so by the
definition of a
XFD $p \rightarrow q$ is violated in $\bar{T}_1$ . This is a contradiction
since $q
\in P^+$ and by Lemma ~\ref{L:term} $\bar{T}_1$ satisfies $p \rightarrow q$ and
so we conclude that $|N(q)| = 1$.
\hfill$\Box$\\

\begin{lemma}
\label{L:b12}
Let $r \rightarrow s$ be a XFD in $\Sigma$, let
$\bar{v}_1.\cdots.\bar{v}_n$ and
$\bar{v}'_1. \cdots. \bar{v}'_n$ be path instances in $Paths(s)$
in $\bar{T_1}$,
and let $x_1 =  \{v| v\in \{v_1, \cdots, v_n\} \wedge v \in N(r \cap s)\}$
and $y_1
= \{v | v \in \{v'_1, \cdots, v'_n\} \wedge v \in N(r \cap s)\}$.  Then if
$p \cap s$ is
a strict prefix of $ r \cap s$ and $p \rightarrow r \notin \Sigma$ and $p
\rightarrow s \notin \Sigma$ then $x_1 \neq y_1$.
\end{lemma}

\begin{figure}[here]
\begin{center}
\includegraphics[scale=.6]{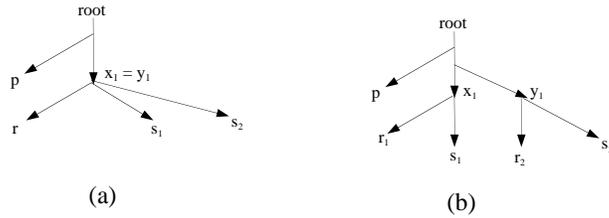}
\end{center}
\caption{A XML tree illustrating Lemma ~\ref{L:b12}}
\label{F:Xstructure}

\end{figure}

{\bf Proof.}   The claim of the lemma can best be illustrated by a diagram.
Let
$s_1$ denote the path instance $\bar{v}_1.\cdots.\bar{v}_n$ and let $s_2$
denote
the path instance $\bar{v}'_1. \cdots. \bar{v}'_n$. Then the claim of the lemma is
that only the situation illustrated in (b) of Figure ~\ref{F:Xstructure}
can arise, and
not the situation illustrated in (a) of Figure ~\ref{F:Xstructure}.  We
prove the
result by induction on the number of steps to generate $\bar{T_1}$.  Firstly we
claim that $T_1$ cannot have the structure illustrated in (a) of Figure
~\ref{F:Xstructure}.  Suppose that it has.  Then since by definition of
$T_1$ there
has to be two path instances for every path and $x_1 = y_1$, there must be
another
distinct node in $N(r \cap s)$.  However, using the same argument as in Lemma
~\ref{L:8} shows that we then contradict the fact that either there are
exactly two
path instances for any path in $T_1$  or we contradict the fact that $T_1$ is
complete.  Hence $T_1$ must have the structure shown in (b) of Figure
~\ref{F:Xstructure}.  Assume inductively then that the property holds after
iteration $k-1$ of Algorithm 1. The only way that (b) of Figure
~\ref{F:Xstructure} could possibly arise is if  we merged path instances of
$r$ or
path instances of $s$ but this cannot occur because of the definition of
Algorithm 1
and the assumptions that $p \rightarrow r \notin \Sigma$ and $p \rightarrow s
\notin \Sigma$.
\hfill$\Box$\\

\begin{theorem}
\label{T:2}

Algorithm 2 correctly computes $P^+$ when $Last(p) \in {\bf E}$.
\end{theorem}

{\bf Proof.} The proof that if $q \in P^+$  then $p \rightarrow q$ is in
$\Sigma^+$
is the same as for Theorem ~\ref{T:1}.

Next we show that if  $p \rightarrow q \in \Sigma^+$ then $q \in P^+$. We
firstly
claim that $\bar{T}_1$ satisfies $\Sigma$. Let $r \rightarrow s$ be any XFD in
$\Sigma$. If $r = root$ then it follows from Lemma ~\ref{L:b4} and Lemma
~\ref{L:term} that $r \rightarrow s$ is satisfied in $\bar{T}_1$. Suppose
then that $r
\rightarrow s$ is violated in $\bar{T}_1$ and $r \neq root$. The first way
for this
to happen is if  there exist two path instances $\bar{v}_1. \cdots.
\bar{v}_n$ and
$\bar{v}'_1. \cdots. \bar{v}'_n$ in $Paths(s)$ such that $x_1 \neq y_1$, where
$x_1 =
\{v| v \in \{\bar{v}_1, \cdots, \bar{v}_n\} \wedge v \in N(r \cap
s) \}$  and
$y_1 =  \{v | v \in \{v'_1, \cdots, \bar{v}'_n\} \wedge v \in
N(r \cap s) \}$, and there exist $v_1 \in Nodes(x_1, p)$ and $v_2 \in
Nodes(y_1,
p)$ such that $val(v_1) = val(v_2)$. Also, since $x_1 \neq y_1$, $v_1$ and
$v_2$
are distinct.   For this to happen it follows from Lemma ~\ref{L:b2} that
$r \in
P^+$.  We must also have that $s \in P^+$ or else $s$ could be added to
$P^+$ at
line 5 thus contradicting the definition of $P^+$.  However, if $s \in P^+$
then by
Lemma ~\ref{L:term} $p \rightarrow s$ is satisfied in $\bar{T}_1$ which
contradicts the assumption  that $r \rightarrow s$ is violated in
$\bar{T}_1$ by
Lemma ~\ref{L:b4}. We conclude that $r \rightarrow s$ is satisfied.  The second
way that $r \rightarrow s$ could be violated in $\bar{T}_1$  is if there
exist two
path instances $\bar{v}_1. \cdots. \bar{v}_n$ and $\bar{v}'_1. \cdots.
\bar{v}'_n$
in $Paths(s)$ such that $x_1 = y_1$. If $x_1 = root$ then the same arguments as
used in Theorem ~\ref{T:1} shows that $r \rightarrow s$ is satisfied. If
$x_1 \neq
root$  then
by Lemma ~\ref{L:b1} there either exists $s'$ such that $s' \in Anc(s)$ and
$s' \in
P^+$ or $s \cap p \neq root$. Consider the first possibility. Using the same
arguments as in Theorem ~\ref{T:1}  shows that $r \rightarrow s$ is satisfied.
Consider then the second situation where $s \cap p \neq root$.  There are
two cases
to consider: (a) $r \cap s = root$ and (b) $r \cap s \neq root$.
Consider (a).  Since $r \rightarrow s \in \Sigma$ and $r \cap s = root$
then $s \in
P^+$ or else it could be added at line 6 thus contradicting the definition
of $P^+$.
So  if $r \rightarrow s$ is violated in $\bar{T}_1$ then by Lemma ~\ref{L:b4}
$p \rightarrow s$ is violated which contradicts  Lemma ~\ref{L:term} and so we
conclude that $r \rightarrow s$ is satisfied in $\bar{T}_1$.
Consider (b).   We now consider the subcases: (b.1) $p \cap s$ is a strict
prefix of
$r \cap s$ and (b.2)  $p \cap s$ is not a strict prefix of $r \cap s$.
Consider (b.1).
Suppose $ p \rightarrow  s \in \Sigma$.  Then $s \in P^+$ or else it could
be added
at line 5 which contradicts the definition of $P^+$.  So by Lemma ~\ref{L:term}
$p \rightarrow s$ is satisfied in $\bar{T}_1$.  Suppose then that $ p
\rightarrow  r
\in \Sigma$. Then $r \in P^+$ or else it could be added at line 5 which
contradicts
the definition of $P^+$ and so $s \in P^+$ or else it could be added at
line 5 which
contradicts the definition of $P^+$.  Assume then that $ p \rightarrow  s
\notin
\Sigma$ and that $ p \rightarrow  r \notin \Sigma$.  Then this case cannot
arise
because by Lemma ~\ref{L:b12} this would imply that $x_1 \neq y_1$ which
contradicts the assumption that $x_1 = y_1$.
Consider (b.2). Since $p \cap s$ is not a strict prefix of $r \cap s$ then
$r \cap s$ is
a prefix of $p$ and so we must have that $r \cap s \in P^+$ or else it
could be added
at line 2 which contradicts the definition of $P^+$. Then since $r \cap s
\in P^+$ it
follows that $s \in P^+$ or else it could be added at line 4 since $r \cap
s$ is a
prefix of $r \cap s$ and $r \cap s$ is a prefix of $r$ and $s$. Hence using
the same
arguments as previously it follows that $r \rightarrow s$ is satisfied.

To complete the proof suppose that  $p \rightarrow q \in \Sigma^+$. Since
$\bar{T}_1$ satisfies $\Sigma$  then $\bar{T}_1$ also satisfies $p
\rightarrow q$.
Suppose firstly that $Last(q) \in {\bf E}$.  If $|N(q)| \neq 1$ then using
a similar
argument to Lemma ~\ref{L:b4}  this would imply that $p \rightarrow q$ is
violated in $\bar{T}_1$ which is a contradiction and so $|N(q)| = 1$. Then by
Lemma ~\ref{L:b3} $q \in P^+$. Suppose instead that that $Last(q) \notin
{\bf E}$.
By definition of Algorithm 1, there exists two nodes in $N(q)$, say $v_1$ and
$v_2$, since $Last(q) \notin {\bf E}$ and non element nodes are not removed in
Algorithm 1. If $val(v_1) \neq val(v_2)$ then using similar arguments to those
used in Lemma  ~\ref{L:b4} it follows that $\bar{T}_1$ violates $p
\rightarrow q$
which is a contradiction. Thus we must have that $val(v_1) = val(v_2)$ and
so by
Lemma ~\ref{L:b2} $q \in P^+$.  This completes the proof.
\hfill$\Box$\\

\section{Conclusions}
In this paper we have investigated issues related to the functional
dependencies in
XML.  Such constraints are important because of the close relationship between
XML and relational databases and also because of the importance of functional
dependencies in developing a theory of normalization.  In an associated paper
\cite{Vin02a} we defined functional dependencies in XML (XFDs) and provided a
set of axioms for reasoning about XFD implication.  In this paper we have
proven
prove that the axioms are also complete for unary XFDs.  The second
contribution
of the paper has been to prove that the implication problem for unary XFDs is
decidable and to provide a linear time algorithm for it.  These results have
considerable significance in the development of a theory of normalization
for XML
documents.  In relational databases, the classic results on soundness and
completeness of Armstrong's axioms \cite{Arm74} and the resulting closure
algorithm for FD implication play an essential role in determining whether a
relation is in one of the classic normal forms.  Similarly, the results in
this paper
are an important first step in the development of algorithms for testing
the normal
form proposed in \cite{Vin02a}.

There are several other issues related to the one investigated in this
paper that we
intend to investigate in the future.  The main results of this paper have
only been
established for unary XFDs and there is a need to extend the results to
arbitrary
XFDs. Secondly, the approach adopted in this paper is based on the strong
satisfaction approach to XFD satisfaction but the techniques we have used can also
be extended to defining weak satisfaction.  In this case there is a need to
a develop
complete and sound axiom system for implication of weak XFDs as well as
determining if the implication of weak satisfaction is decidable and if so
to develop
an efficient algorithm for it.  Thirdly, there is a need to investigate the
extension of
the other important class of constraints in relational databases, namely
multivalued
dependencies (MVDs) \cite{Fag77}, to XML  We have already completed some
preliminary work on this problem \cite{Vin02b} and in particular we have
defined
MVDs in XML and proposed a 4NF for XML and shown that it eliminates
redundancy. However, important issues such as axiom systems for MVDs and the
interaction between XFDs and MVDs in XML have yet to be investigated.

\bibliographystyle{plain}
\bibliography{temp}

\section{Appendix}

-------
{\bf Proof of Theorem ~\ref{T:sound}}

Axiom A1 is immediate from the definition of a XFD.

Consider A2. Suppose that there exists two distinct path instances $\bar{v}_1.
\cdots. \bar{v}_n$ and
$\bar{v}'_1. \cdots. \bar{v}'_n$ in
$Paths(q)$ satisfying the conditions in Definition  ~\ref{D:Xfd2}.   Then by
Definition ~\ref{D:Xfd2} we must have that $\exists i, 1 \leq i \leq k$,
such that
$x_i \neq y_i$ if $Last(p_i) \in {\bf E}$ else
$\exists i, 1 \leq i \leq k$, such that $\perp \notin Nodes(x_i, p_i) $ and
$\perp
\notin Nodes(y_i, p_i) $ and $
val(Nodes(x_i,p_i)) \cap val(Nodes (y_i, p_i)) = \phi$.  This condition
will still
hold for the XFD $p, p_1, \cdots, p_k
\rightarrow q$ and so A2 is sound.

Consider A3. Suppose that there exists two distinct path instances $\bar{v}_1.
\cdots. \bar{v}_n$ and
$\bar{v}'_1. \cdots. \bar{v}'_n$ in
$Paths(s)$ such that. Then since $q \rightarrow s$ is satisfied, we must
have that $
x'_1 \neq y'_1$ where $x'_1 =
\{v |
v \in \{v_1, \cdots, \bar{v}_n \} \wedge v \in N(q \cap s)\}$  and
$y'_1 =
\{v | v \in \{v'_1, \cdots, \bar{v}'_n\} \wedge v
\in N(q \cap
s)\}$.. Then there must exist path instances $\bar{t}_1. \cdots. \bar{t}_n$ and
$\bar{t}'_1. \cdots. \bar{t}'_n$  in $Paths(q)$ such that $x'_1$ is in
$\bar{t}_1.
\cdots. \bar{t}_n$ and $y'_1$ is in $\bar{t}'_1. \cdots. \bar{t}'_n$.
Since $x'_1
\neq y'_1$ the two path instances must be distinct. Also, since $q
\rightarrow s$ is
satisfied we must have that $val(\bar{t}_n) \neq val(\bar{t}'_n)$.  So by the
definition of a XFD and since $p_1, \cdots, p_k \rightarrow q$ is satisfied we
must have that $\exists i, 1 \leq i \leq k$, such that $x_i \neq y_i$
(where $x_i$ and
$y_i$ are defined as in Definition ~\ref{D:Xfd2}) if $Last(p_i) \in {\bf
E}$ else
$\exists i, 1 \leq i \leq k$, such that $\perp \notin Nodes(x_i, p_i) $ and
$\perp
\notin Nodes(y_i, p_i) $ and $
val(Nodes(x_i,p_i)) \cap val(Nodes (y_i, p_i)) = \phi$ and so $p_1, \cdots, p_k
\rightarrow s$ is satisfied.

Consider A4. Suppose that there exists two distinct path instances $\bar{v}_1.
\cdots. \bar{v}_n$ and
$\bar{v}'_1. \cdots. \bar{v}'_n$ in
$Paths(q)$ satisfying the conditions (($\bar{v}_n = \perp \wedge
\bar{v}'_n = \perp ) \vee (\bar{v}_n \neq \perp \wedge \bar{v}'_n = \perp )
\vee
(\bar{v}_n = \perp \wedge
\bar{v}'_n
\neq \perp ) \vee (\bar{v}_n \neq \perp \wedge \bar{v}'_n \neq \perp \wedge
val(\bar{v}_n) \neq
val(\bar{v}'_n))$.  Then since $p_i \cap q = root$, it follows that $x_i =
y_i =
root$ for all $i$, $1 \leq i \leq k$ and so $p_1, \cdots, p_k \rightarrow q$ is
violated which is a contradiction. Hence there cannot exist distinct path
instances
$\bar{v}_1. \cdots. \bar{v}_n$ and
$\bar{v}'_1. \cdots. \bar{v}'_n$ in
$Paths(q)$ satisfying the conditions (($\bar{v}_n = \perp \wedge
\bar{v}'_n = \perp ) \vee (\bar{v}_n \neq \perp \wedge \bar{v}'_n = \perp )
\vee
(\bar{v}_n = \perp \wedge
\bar{v}'_n
\neq \perp ) \vee (\bar{v}_n \neq \perp \wedge \bar{v}'_n \neq \perp \wedge
val(\bar{v}_n) \neq
val(\bar{v}'_n))$ and so any XFD $p'_1, \cdots, p'_j \rightarrow q$ is
automatically satisfied.

Consider A5. Suppose firstly that $p \cap q$ is prefix of $p'$ and $p'$ is
a prefix
of $p$  and that $p' \rightarrow q$ is violated. Then since  $p'$ is a
prefix of $p$
we can ignore the case where $p' = p$ and assume that $Last(p') \in {\bf
E}$.  So
for $p' \rightarrow q$ to be violated we must have that $x'_1 = y'_1$ where
$x'_1
=
\{v |
v \in \{v_1, \cdots, \bar{v}_n \} \wedge v \in N(p' \cap q)\}$  and
$y'_1 =
 \{v | v \in \{v'_1, \cdots, \bar{v}'_n\}
\wedge v \in N(p' \cap
q)\}$.  However since $p'$ is a prefix of $p$ and $p \cap q$ is prefix of
$p'$, it
follows that $x_1 = x'_1$ and $y_1 = y'_1$, where $x_1 =
\{ |
v \in \{v_1, \cdots, \bar{v}_n \} \wedge v \in N(p \cap q)\}$  and
$y_1 =
\{v | v \in \{v'_1, \cdots, \bar{v}'_n\} \wedge v
\in N(p \cap
q)\}$.  Thus it follows that $x_1 = y_1$ and so $Nodes(x_1, p) = Nodes(y_1, p)$
and so $p \rightarrow q$ is violated which is a contradiction and so $p'
\rightarrow
q$ is satisfied. Next suppose that $p \cap q$ is prefix of $p'$ and $p'$ is
a prefix of
$q$  and that $p' \rightarrow q$ is violated. Then since  $p'$ is a prefix
of $p$ we
can ignore the case where $p' = q$ and assume that $Last(p') \in {\bf E}$.
So for
$p' \rightarrow q$ to be violated we must have that $x'_1 = y'_1$ where $x'_1 =
\{v|
v \in \{v_1, \cdots, \bar{v}_n \} \wedge v \in N(p' \cap q)\}$  and
$y'_1 =
\{v |v \in \{v'_1, \cdots, \bar{v}'_n\} \wedge v
\in N(p' \cap
q)\}$.  Then since $p \cap q$ is a prefix of $ p'$ this implies that $x_1 =
y_1$
which implies a contradiction as before.

Consider A6. Suppose that there exists two distinct path instances $\bar{v}_1.
\cdots. \bar{v}_n$ and
$\bar{v}'_1. \cdots. \bar{v}'_n$ in $Paths(q)$ such that $val(\bar{v}_n) \neq
val(\bar{v}'_n )$. Then because $q$ is a prefix of $p$, $p \cap q = q$ and
so $x_1
= \bar{v}_n$ and $y_1 =  \bar{v}'_n$.  Thus $p \rightarrow q$ is satisfied
since
$val(\bar{v}_n) \neq val(\bar{v}'_n)$

Consider A7. Suppose that there exists two distinct path instances $\bar{v}_1.
\cdots. \bar{v}_n$ and
$\bar{v}'_1. \cdots. \bar{v}'_n$ in $Paths(q)$ such that $ val(\bar{v}_n) \neq
val(\bar{v}'_n)$.   Then because $Last(q) \in {\bf A}$ $Parent(\bar{v}_n) \neq
Parent(\bar{v}'_n)$.
Also, by definition of $x_1$ and $y_1$, $x_1 = Parent(\bar{v}_n)$ and $y_1 =
Parent(\bar{v}'_n)$ and thus $x_1 \neq y_1$ and so $Parnt(q) \rightarrow q$ is
Satisfied since $Last(Parnt(q)) \in {\bf E}$.

Axiom A8 is automatic since there is only one path instance that ends with
$v_r$.
\hfill$\Box$\\

--------
{\bf Proof of Theorem ~\ref{T:complete}}

Let $\Sigma$ be a set of XFDs and let $\Sigma^+$ be the set of XFDs
obtained
by using Axioms A1 - A8.  Let $p \rightarrow q$ be a XFD that is not in
$\Sigma^+$.
Then to show completeness it suffices to show that there exists a tree $T$ that
satisfies $\Sigma$ but not $p \rightarrow q$.  We consider several cases.

\underline{Case A: $Last(p) \in {\bf E}$}

We now consider several subcases. The only cases that can arise are: (a) $
p > q$; (b) $q > p$; (c) $p \not> q$ and $q \not> p$. We
firstly note that because of Axiom A6 case (a) cannot arise so the only
cases to
consider are (b) and (c).  We consider (c) first.

\underline{Case AA: $p \not> q$ and $q \not> p$}

Let $\{p_1 \rightarrow q, \ldots, p_n
\rightarrow q\}$ be the set of all XFDs in $\Sigma^+$ which have $q$ on the
r.h.s
(we note that $\Sigma^+$ can be computed using Algorithm 2  in Section 5).
Consider the paths $\{p_1 \cap q, \ldots, p_n \cap q\}$.  Since each of
these paths is
a prefix of  $q$ we can order the set $\{p_1 \cap q, \ldots, p_n \cap q\}$
according to $>$.  Let $p_{min}$ be the minimum of
$\{p_1 \cap q, \ldots, p_n \cap q\}$.  We firstly claim that $p_{min} \neq
root$.  If
not, then there exists $p_i \rightarrow q$ such that $ p_i \cap q = root$
and so by
A4 $p \rightarrow q \in \Sigma^+$ which is a contradiction.  Next we claim that
$p_{min} \rightarrow q$.  This follows from the definition of $p_{min}$ and
axiom A5. Define the  node $p_{branch}$ by $p_{branch} = Parnt(p_{min})$.

Construct then a tree $T$ with the following properties. $T$ is complete w.r.t.
$P_\Sigma$.  For all paths $p'$ such that $p' \cap p_{branch}$ is a strict
prefix of
$p_{branch}$, $T$ contains one path instance for $p'$. If $p' \cap
p_{branch}$ is
not a strict prefix of $p_{branch}$ then $T$ contains exactly two path
instances for
$p'$. Moreover, if $Last(p') \notin {\bf E}$ then the $val$ of the two nodes in
$N(p')$ are distinct if $p' \rightarrow q \in \Sigma^+$ otherwise they are
the same.
Such a tree always exists.  It is also clear from this construction that
$T$ violates $p
\rightarrow q$. We also note that $T$ has the property that $p_{min}$ (and
hence $p_{branch}$) cannot
be a prefix of $p$.  If it was then $p \rightarrow p_{min}$ by A6 and since
$p_{min} \rightarrow q$ then by A3 $p \rightarrow q \in \Sigma^+$ which is a
contradiction.

We illustrate the construction by an example. Let $p \rightarrow q$ be the XFD
\texttt{root.X} $\rightarrow$ \texttt{root.A.B.C.D.E.E\#} and let $\Sigma =
\{\texttt{root.A.B.C.C\#} \rightarrow \texttt{root.A.B.C.D.E.E\#},
\texttt{root.A.B.C.D.D\#} \rightarrow \texttt{ root.A.B.C.D.E.E\#},$

\begin{figure}[here]
\begin{center}
\includegraphics[scale=.6]{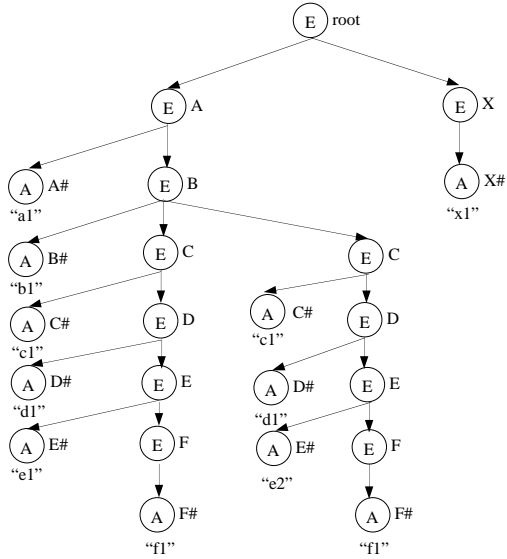}
\end{center}
\caption{A XML tree }
\label{F: Tcomplete}
\end{figure}

$\texttt{root.X.X\#} \rightarrow \texttt{root.A}\}$. Then the above
construction
procedure yields the tree $T$ shown in Figure ~\ref{F: Tcomplete}.

We note that in the construction of $T$,  the correct place to branch the
tree is
crucial. If the tree branches above or below $p_{branch}$ then $T$
will not satisfy $\Sigma$.

\begin{figure}[here]
\begin{center}
\includegraphics[scale=.6]{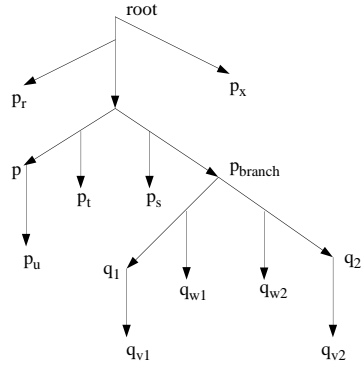}
\end{center}
\caption{A XML tree }
\label{F: Tproof}
\end{figure}

It is clear from this construction that $T$ violates $p \rightarrow q$ so
it remains
to prove that $T$ satisfies $\Sigma$.  Let $ p' \rightarrow q'$ be any XFD
(not necessarily in $\Sigma$).  There are several cases to consider
depending on where $ p'$ and $q'$
are in the tree $T$.  The different cases can be best illustrated by Figure
~\ref{F: Tproof}. In this figure we use subscripts to denote different
instances of a
path.  For example, $q_1$ and $q_2$ denote different path instances of the path
$q$ and $q_{v1}$ and $q_{v2}$ denote different path instances of the path $q_v$.
We shall consider all possible cases where $ p' \rightarrow q'$  could be
violated in $T$ and show that either $ p' \rightarrow q'$ cannot be in
$\Sigma$ or $T$ satisfies $
p' \rightarrow q'$.

\underline{Case AAA: $ p' = p_x$, i.e. $ p' \cap q = root$}

\underline{Case AAA.1: $q_w > q' \geq p_{min} $ }

Suppose that $ p' \rightarrow q' \in \Sigma$. Since $q' \geq p_{min}$ and
$q' \in {\bf E}$ (since $q_w > q'$ ) it follows from A6 or A1 that
$ q' \rightarrow p_{min}$. Then
since $p_{min} \rightarrow q$ and applying A3 twice we derive that $p'
\rightarrow q$. Then since $ p' \cap q = root$ by A4 we derive that $p
\rightarrow
q \in \Sigma^+$  which  is a contradiction and so we conclude that $ p'
\rightarrow
q' \notin \Sigma$.

\underline{Case AAA.2: $q' = q_w$, i.e. $q > q' \cap q \geq p_{min}$ and
$Last(q') \notin {\bf E}$}

Suppose that $ p' \rightarrow q' \in \Sigma$  and $ p' \rightarrow q' $ is
violated in $T$. Then for this to happen the two nodes in $N(q')$ must have
different $val's$  and so by the construction of $T$, $ q'
\rightarrow q \in \Sigma^+$ and so by A3 $ p' \rightarrow q$.  By A4 this
implies that $p \rightarrow q \in \Sigma^+$ which is a contradiction and so we
conclude that either $ p' \rightarrow q'$ is satisfied in $T$ or $ p'
\rightarrow q' \notin \Sigma$   .

\underline{Case AAA.3: $q' = q$}

 If  $ p'  \rightarrow q' \in \Sigma $ then by A4 $p \rightarrow q \in \Sigma^+
$ which is a contradiction and so we conclude that $ p'  \rightarrow q' \notin
\Sigma $.

\underline{Case AAA.4: $ q' > q$ and $Last(q') \in {\bf E}$ }

Suppose $p' \rightarrow q' \in \Sigma $. Since $q' > q$ and $q > p_{min}$, then $q'
\rightarrow  p_{min} \in  \Sigma ^+ $ by A6 and since $p_{min}
\rightarrow q$ by A3 this implies that
$ q' \rightarrow q$.  So if $p' \rightarrow q'$ then by A3 $p' \rightarrow
q$ so by
A4 $p \rightarrow q \in \Sigma^+$ which is a contradiction and so $p'
\rightarrow
q' \notin \Sigma $.

\underline{Case AAA.5: $q' = q_v$, i.e. $ q' > q$ and $Last(q') \notin {\bf
E}$ }

As for Case AAA.2
\\

\underline{Case AAB: $ p_x  > p' $ }

As  for Case AAA.
\\

\underline{Case AAC: $ p' = p_r $, i.e. $ p \cap q \geq p' \cap q > root$ }

\underline{Case AAC.1: $q' = q$}

If $ p' \rightarrow q' \in \Sigma$ then it contradicts the definition of
$p_{min}$
and so $ p' \rightarrow q' \notin \Sigma$.

\underline{Case AAC.2: $q_w > q' \geq p_{min} $ }

Suppose that $ p' \rightarrow q' \in \Sigma$. Since $q' \geq p_{min}$ and
$q' \in {\bf E}$ it follows from A6 that
$ q' \rightarrow p_{min}$. Then since $p_{min} \rightarrow q$ and
applying A3 twice we derive that $ p'
\rightarrow q \in \Sigma^+$ which contradicts the definition of $p_{min}$
and so $
p' \rightarrow q' \notin \Sigma$.

\underline{Case AAC.3: $q' = q_w$, i.e. $q > q' \cap q \geq p_{min}$ and
$Last(q') \notin {\bf E}$ }

Suppose that $ p' \rightarrow q' \in \Sigma$  and $ p' \rightarrow q' $ is
violated in $T$. Then for this to happen the two nodes in $N(q')$ must have
different $val's$  and so by the construction of $T$, $q'
\rightarrow q \in \Sigma^+$ and so by A3 $ p' \rightarrow q$.  By A4 this
implies that $p \rightarrow q \in \Sigma^+$ which contradicts the
definition of $p_{min}$ and so we
conclude that either $ p' \rightarrow q'$ is satisfied in $T$ or $ p'
\rightarrow q' \notin \Sigma$.

\underline{Case AAC.4: $ q' > q$ and $Last(q') \in {\bf E}$ }

Suppose that $ p' \rightarrow q' \in \Sigma$. Since $q' > q$ and $q >
p_{min}$, then $q'
\rightarrow  p_{min} \in  \Sigma ^+ $ and since $p_{min} \rightarrow q$ it
follows by A3 that $ p'
\rightarrow q$  which
contradicts the definition of $p_{min}$ and so $ p' \rightarrow q' \notin
\Sigma$.

\underline{Case AAC.5: $q' = q_v$, i.e. $ q' > q$ and $Last(q') \notin {\bf
E}$ }

If $p' \rightarrow q'$ is violated in $T$, then it follows by construction
of $T$ that
$q'
\rightarrow q$ and so by A3 $p' \rightarrow q$ which contradicts the
definition of
$p_{min}$ and so $p' \rightarrow q'$ is satisfied.
\\

\underline{Case AAD: $ p_r  > p' $ }

\underline{Case AAD.1: $q' = q$}

Assume that $ p' \rightarrow q' \in \Sigma$.  If $p'$ is a prefix of $p$
then by A6
$p \rightarrow p'$ and so since $p'
\rightarrow q$ it follows by A3 that $p \rightarrow q \in \Sigma^+$ which is a
contradiction and so $ p' \rightarrow q' \notin \Sigma$.  If $p'$ is not a
prefix of
$p$ then it follows that from the fact that
$p' \rightarrow q$ and A5 that
$ p' \cap p \rightarrow q$  and since $p \rightarrow p' \cap p$ by A6 then
applying A3 derives the contradiction that $p \rightarrow q \in \Sigma^+$
and so $ p' \rightarrow q' \notin \Sigma$.

\underline{Case AAD.2: $q_w > q' \geq p_{min} $ }

Assume that $ p' \rightarrow q' \in \Sigma$.  Then $ p' \cap q \rightarrow
q$ by
A4.  However by definition of $ p'$, $ p' \cap q$ is a prefix of $p$ and so
by A6
$p \rightarrow p' \cap q$ and so by A3 $p \rightarrow q$ which is a
contradiction
and so $ p' \rightarrow q' \notin \Sigma$.

\underline{Case AAD.3: $q' = q_w$, i.e. $q > q' \cap q \geq p_{min} $ and
$Last(q') \notin {\bf E}$  }

Suppose that $ p' \rightarrow q' \in \Sigma$  and $ p' \rightarrow q' $ is
violated in $T$. Then for this to happen the two nodes in $N(q')$ must have
different $val's$  and so by the construction of $T$, $q'
\rightarrow q \in \Sigma^+$ and so by A3 $ p' \rightarrow q$. However using
the same argument as in AAD.2, if $
p' \rightarrow q$ then $p \rightarrow q \in \Sigma^+$ which is a
contradiction and so we
conclude that either $ p' \rightarrow q'$ is satisfied in $T$ or $ p'
\rightarrow q' \notin \Sigma$   .

\underline{Case AAD.4: $ q' > q$ and $Last(q') \in {\bf E}$ }

Assume that $ p' \rightarrow q' \in \Sigma$.  As in AAC.4 we derive that $ p'
\rightarrow q$ and so using the same argument as in AAD.2 we derive the
contradiction that $p \rightarrow q \in \Sigma^+$ and so $ p' \rightarrow
q' \notin
\Sigma$.

\underline{Case AAD.5: $q' = q_v$, i.e. $ q' > q$ and $Last(q') \notin {\bf
E}$ }

As in AAC.5 we derive that $ p' \rightarrow q$ and so using the same
argument as
in AAD.2 we derive the contradiction that $p \rightarrow q$.
\\

\underline{Case AAE: $p' = p_s$, i.e. $p_{min} > p' \cap q > p \cap q$ }

As for case for Case AAC.
\\

\underline{Case AAF: $ p_s  > p' $ }

We can assume that $p \cap q$ is a prefix of $ p'$ or else the case reduces
to  case
AAD.

\underline{Case AAF.1: $q' = q$}

Assume that $ p' \rightarrow q' \in \Sigma$.  Since $p' \rightarrow q$, by
A5 we
have that $p' \cap q \rightarrow q$ and since $ p_s  > p' $ it follows that
$p' \cap q$ is a strict prefix of $p_{min}$ which contradicts the definition of
$p_{min}$ and so $ p' \rightarrow q' \notin \Sigma$.

\underline{Case AAF.2: $q_w > q' \geq p_{min} $  }

Suppose that $p' \rightarrow q' \in \Sigma$. Since $p_{min} \rightarrow q$ it
follows from A5 and the definition of $ q'$ that $ q' \cap q \rightarrow q$.
However since $q' \geq p_{min}$ then $Last(q') \in {\bf E}$ and so since $
q' \cap
q$ is prefix of $q'$ it follows that $ q' \rightarrow q' \cap q$.  So
applying A3
twice we derive that $p' \rightarrow q$ which contradicts the definition of
$p_{min}$ and so $p' \rightarrow q' \notin \Sigma$.

\underline{Case AAF.3: $q' = q_w$, i.e. $q > q' \cap q \geq p_{min}$ and
$Last(q') \notin {\bf E}$   }

Suppose that $ p' \rightarrow q' \in \Sigma$  and $ p' \rightarrow q' $ is
violated in $T$. Then for this to happen the two nodes in $N(q')$ must have
different $val's$  and so by the construction of $T$, $ q'
\rightarrow q \in \Sigma^+$ and so by A3 $ p' \rightarrow q \in \Sigma ^+$
which again contradicts the definition of
$p_{min}$. So we
conclude that either $ p' \rightarrow q'$ is satisfied in $T$ or $ p'
\rightarrow q' \notin \Sigma$   .

\underline{Case AAF.4: $ q' > q$ and $Last(q') \in {\bf E}$ }

Suppose that $p' \rightarrow q' \in \Sigma$. Since $q' > q$ and $q >
p_{min}$, then $q'
\rightarrow  p_{min} \in  \Sigma ^+ $ by A6. Then since
$p_{min} \rightarrow q$, it follows by applying A3 three times that $p'
\rightarrow q \in \Sigma^+$ which contradicts the definition of $p_{min}$
and so
$p' \rightarrow q' \notin \Sigma$.

\underline{Case AAF.5: $q' = q_v$, i.e. $ q' > q$ and $Last(q') \notin {\bf E}$}

Suppose that $ p' \rightarrow q' \in \Sigma$  and $ p' \rightarrow q' $ is
violated in $T$. If $p' \rightarrow q'$ is violated in $T$ then by
definition of $T$ we must have
that $q' \rightarrow q$  and so by A3 $p' \rightarrow q$ which contradicts the
definition of $p_{min}$. So we
conclude that either $ p' \rightarrow q'$ is satisfied in $T$ or $ p'
\rightarrow q' \notin \Sigma$.\\

\underline{Case AAG: $ p' = p_t$, i.e. $p > p' \cap q > p \cap q$ }

\underline{Case AAG.1: $q' = q$}

If $p' \rightarrow q \in \Sigma$ then by A5 $p' \cap q \rightarrow q \in
\Sigma^+$
and by A6
$p \rightarrow p' \cap q$ and so by A3 $p \rightarrow q \in \Sigma^+$ which
is a
contradiction.and so $p' \rightarrow q \notin \Sigma$.

\underline{Case AAG.2: $q_w > q' \geq p_{min} $ }

If $p' \rightarrow q' \in \Sigma$, then by A6 $ q' \rightarrow p_{min}$ and
since $p_{min} \rightarrow q$ it follows by applying A3
twice that $p' \rightarrow q$. Then by A5 $p' \cap q \rightarrow q$ and
since $p
\rightarrow p' \cap q$ by A6 we derive the contradiction $p \rightarrow q \in
\Sigma^+$ and so $p' \rightarrow q' \notin \Sigma$.

\underline{Case AAG.3: $q' = q_w$, i.e. $q > q' \cap q \geq p_{min}$ and
$Last(q') \notin {\bf E}$   }

Suppose that $ p' \rightarrow q' \in \Sigma$  and $ p' \rightarrow q' $ is
violated in $T$. Then for this to happen the two nodes in $N(q')$ must have
different $val's$  and so by the construction of $T$, $ q'
\rightarrow q \in \Sigma^+$ and so by A3 $ p' \rightarrow q \in \Sigma ^+$
which leads to a contradictionas in Case AAD. So we
conclude that either $ p' \rightarrow q'$ is satisfied in $T$ or $ p'
\rightarrow q' \notin \Sigma$   .

\underline{Case AAG.4: $ q' > q$ and $Last(q') \in {\bf E}$ }

Using the same reasoning as in Case AAG.2, if $p' \rightarrow q' \in
\Sigma$,  then we derive the contradiction $p
\rightarrow q \in \Sigma^+$ and so $p' \rightarrow q' \notin \Sigma$.

\underline{Case AAG.5: $q' = q_v$, i.e. $ q' > q$ and $Last(q') \notin {\bf
E}$ }

Suppose that $ p' \rightarrow q' \in \Sigma$  and $ p' \rightarrow q' $ is
violated in $T$. Then by the construction of $T$,
for this to happen we must have that $q' \rightarrow q$ and so by A3 $p'
\rightarrow q$. Then using the same reasoning as in AAG.2 we derive the
contradiction $p \rightarrow q \in \Sigma^+$. So we
conclude that either $ p' \rightarrow q'$ is satisfied in $T$ or $ p'
\rightarrow q' \notin \Sigma$.\\

\underline{Case AAH: $ p_t  > p' \geq p_{min}$ }

 As for case AAG.
\\

\underline{Case AAI: $ p' = p_u$, i.e. $ p' > p$ and $Last(p') \notin {\bf E}$}

\underline{Case AAI.1: $q' = q$}

If $p' \rightarrow q \in \Sigma$ then since $p$ is a prefix of $p'$ it
follows by A5 that $p
\rightarrow q \in \Sigma^+$ which is a contradiction and so $p' \rightarrow
q \notin \Sigma$.

\underline{Case AAI.2: $q_w > q' \geq p_{min} $ }

Same as Case AAG.2.

\underline{Case AAI.3: $q' = q_w$, i.e. $q > q' \cap q \geq p_{min} $ and
$Last(q') \notin {\bf E}$ }

As for Case AAG.3.

\underline{Case AAI.4: $ q' > q$ and $Last(q') \in {\bf E}$ }

As for AAG.2.

\underline{Case AAI.5: $q' = q_v$, i.e. $ q' > q$ and $Last(q') \notin {\bf
E}$ }

As for case AAG.5.\\

\underline{Case AAJ: $ p_u  > p' > p $}.

As for Case AAI.\\

\underline{Case AAK: $ p' = q_w$, i.e. $q > p' \cap q \geq p_{min} $ and
$Last(p') \notin {\bf E}$ }

\underline{Case AAK.1: $q' = q$}

By the construction of $T$, if $ p' \rightarrow q' \in \Sigma$  then the
two nodes
in $N(p')$ have distinct $val$  and so  $ p' \rightarrow q'$ is satisfied
in $T$.

\underline{Case AAK.2: $q_w > q' \geq p_{min} $ }

By A6 it follows that $q' \rightarrow p_{min}$ and since $p_{min}
\rightarrow q$ it follows by A3 that $q' \rightarrow q$.  So if $ p'
\rightarrow q'$ then by A3 we have that $ p' \rightarrow q$. Hence by the
construction of $T$ the two nodes in $N(p')$ must have different $val's$
and so by
the definition of a XFD
$ p' \rightarrow q'$ must be satisfied in $T$.

\underline{Case AAK.3: $ q' > q$ and $Last(q') \in {\bf E}$ }

As for case AAG.2 it follows that $q' \rightarrow q$, and so from A3 $p'
\rightarrow q$ which, using the same reasoning as in case AAK.2, implies
that $ p' \rightarrow q'$ is satisfied in
$T$.

\underline{Case AAK.4: $q' = q_v$, i.e. $ q' > q$ and $Last(q') \notin {\bf
E}$ }

Suppose that $p' \rightarrow q' \in \Sigma$ and $p' \rightarrow q'$ is
violated in $T$.  For this to happen we must
have the two nodes in $N(p')$ have the same $val$ and the two nodes in $N(q')$
have different $val's$.  However, by the construction of $T$ if the two
nodes in
$N(q')$ have different $val's$ then $q' \rightarrow q \in \Sigma^+$.  So
applying
A3 we derive that $p' \rightarrow q \in \Sigma^+$ and by the definition of
$T$ this
implies that the two nodes in $N(p')$ must have different $val's$ which is a
contradiction. So $p' \rightarrow q' \notin \Sigma$ or $p' \rightarrow q'$
is satisfied in $T$.
\\

\underline{Case AAL: $ p' = q_v$, i.e. i.e. $ q' > q$ and $Last(q') \notin
{\bf E}$ }

\underline{Case AAL.1: $q' = q$}

As for case AAK.1.

\underline{Case AAL.2: $q_w > q' \geq p_{min} $ }

As for case AAK.2

\underline{Case AAL.3: $ q' > q$ and $Last(q') \in {\bf E}$ }

As for case AAK.3\\

\underline{Case AB: $q > p$}

We firstly note that because of axiom A7 we can rule out the case where $q \in
Att(p)$.
Let $\{p_1 \rightarrow q, \ldots, p_n
\rightarrow q\}$ be the set of all XFDs in $\Sigma^+$ which have $q$ on the
r.h.s
(we note that $\Sigma^+$ can be computed using Algorithm 2  in Section 6).
Consider the paths $\{p_1 \cap q, \ldots, p_n \cap q\}$.  Since each of
these paths is
a prefix of  $q$ we can order the set $\{p_1 \cap q, \ldots, p_n \cap q\}$
according to $>$.  Let $p_{min1}$ be the minimum of
$\{p_1 \cap q, \ldots, p_n \cap q\}$ such that $p_{min1} > p$.  We note that
$p_{min1}$ and $p$ are comparable since both are prefixes of $q$.  Define the
node $p_{branch1}$
by $p_{branch1} = Parnt(p_{min1})$.  We also note that since $p_i
\rightarrow q$, it follows from axiom A5 that $p_{min1} \rightarrow q$.

Construct then a tree $T$ with the following properties $T$ is complete w.r.t.
$P_\Sigma$.  For all paths $p'$ such that $p' \cap p_{branch1}$ is a strict
prefix of
$p_{branch1}$, $T$ contains one path instance for $p'$. If $p' \cap
p_{branch1}$ is
not a strict prefix of $p_{branch1}$ then $T$ contains two path instances
for $p'$.
Moreover, if $Last(p') \notin {\bf E}$ then the $val$ of the two nodes in
$N(p')$
are distinct if $p' \rightarrow q \in \Sigma^+$ otherwise they are the
same.  Such a
tree always exists.  It is also clear from this construction that $T$
violates $p
\rightarrow q$. As before, we note that the decision of where to branch the
tree is
critical in the construction of a tree which satisfies $\Sigma$.  We claim
that $T$
satisfies $\Sigma$. As before we let $ p' \rightarrow q'$ be any XFD (not
necessarily in $\Sigma$).
The various cases that can arise are illustrated in Figure ~\ref{F:
Tproof1}. In this figure, as previously, we use subscripts to denote
different instances of a
path.
We shall consider all possible cases where $ p' \rightarrow q'$  could be
violated in $T$
and show that either $ p' \rightarrow q'$ cannot be in $\Sigma$ or $T$
satisfies $
p' \rightarrow q'$.

\begin{figure}[here]
\begin{center}
\includegraphics[scale=.6]{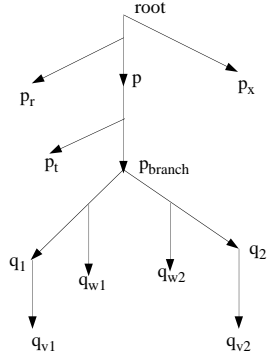}
\end{center}
\caption{A XML tree }
\label{F: Tproof1}
\end{figure}

\underline{Case ABA: $ p' = p_x$, i.e. $ p' \cap q = root$}

\underline{Case ABA.1: $q_w > q' \geq p_{min1} $ }

Suppose that $p' \rightarrow q' \in \Sigma $. Then by A6 $q' \rightarrow
p_{min1}$ and $p_{min1} \rightarrow q$ and so by A3 $p' \rightarrow q \in
\Sigma^+$.  Then by A4 $p \rightarrow q \in \Sigma^+$ which is a contradiction
and so $p' \rightarrow q' \notin \Sigma$.

\underline{Case ABA.2: $q' = q_w$, i.e. $q > q' \cap q \geq p_{min}$ and
$Last(q') \notin {\bf E}$  }

As for case AAA.2.

\underline{Case ABA.3: $q' = q$}

As for case AAA.3.

\underline{Case ABA.4: $ q' > q$ and $Last(q') \in {\bf E}$ }

Suppose that $p' \rightarrow q' \in \Sigma$. Since $q' > q$ and $q >
p_{min1}$, then $q'
\rightarrow  p_{min1} \in  \Sigma ^+ $ by A6 and since by definition
$p_{min1} \rightarrow q$ by A3 this implies that
$ q' \rightarrow q$. As for case ABA.1  this implies $p \rightarrow q \in
\Sigma^+$ which is a contradiction and so $p' \rightarrow q' \notin \Sigma$.

\underline{Case ABA.5: $q' = q_v$, i.e. $ q' > q$ and $Last(q') \notin {\bf
E}$ }

As for Case AAA.2
\\

\underline{Case ABB: $ p_x  > p' $ }

As  for Case AAA.
\\

\underline{Case ABC: $ p' = p_r $, i.e. $ p \cap q \geq p' \cap q > root $}

\underline{Case ABC.1: $q' = q$}

If $ p' \rightarrow q \in \Sigma$ then by A5 $p \rightarrow q \in \Sigma^+$
which is a contradiction and so $ p' \rightarrow q \notin \Sigma$.

\underline{Case ABC.2: $q_w > q' \geq p_{min1} $ }

Suppose that $p'  \rightarrow q' \in \Sigma$. Since $ q' \geq p_{min1}  $ and
since $q' \in {\bf E}$ it follows from A6 that
$ q' \rightarrow p_{min1}$. Also since $p_{min1} \rightarrow q$ by applying
A3 twice we derive that $p'
\rightarrow q$ and so by A5 $p \rightarrow q \in \Sigma^+$ which is a
contradiction and so $p'  \rightarrow q' \notin \Sigma$.

\underline{Case ABC.3: $q' = q_w$, i.e. $q > q' \cap q \geq p_{min}$ and
$Last(q') \notin {\bf E}$   }

Suppose that $ p' \rightarrow q' \in \Sigma$  and $ p' \rightarrow q' $ is
violated in $T$. Then for this to happen the two nodes in $N(q')$ must have
different $val's$  and so by the construction of $T$, $ q'
\rightarrow q \in \Sigma^+$ and so by A3 $ p' \rightarrow q \in \Sigma^+$
which, as for case ABC.2, is a
contradiction. So we 
conclude that either $ p' \rightarrow q'$ is satisfied in $T$ or $ p'
\rightarrow q' \notin \Sigma$   .

\underline{Case ABC.4: $ q' > q$ and $Last(q') \in {\bf E}$ }

Suppose that $ p' \rightarrow q' \in \Sigma$  and $ p' \rightarrow q' $ is
violated in $T$. Since $ q' > q$  and $q > p_{min1}$ then from A6 $q'
\rightarrow p_{min1}$ and since
$p_{min1} \rightarrow q$ it follows by A3 that $ p' \rightarrow q$  which,
as in
case ABC.2, is a contradiction. So $p' \rightarrow q'$ is satisfied in $T$
or $ p' \rightarrow q' \notin \Sigma$.

\underline{Case ABC.5: $q' = q_v$, i.e. $ q' > q$ and $Last(q') \notin {\bf
E}$ }

As for case ABC.3
\\

\underline{Case ABD: $ p' = p_t$, i.e. $p > p' \cap q > p \cap q$ }

\underline{Case ABD.1: $q' = q$}

$p' \rightarrow q'$ cannot be in $\Sigma$ or else it contradicts the
definition of
$p_{min1}$.

\underline{Case ABD.2: $q_w > q' \geq p_{min1} $ }

If $p' \rightarrow q' \in \Sigma$, then by the same reasoning as in AAG.2  $p'
\rightarrow q$ which contradicts the definition of $p_{min1}$.

\underline{Case ABD.3: $q' = q_w$, i.e. $q > q' \cap q \geq p_{min} $ and
$Last(q') \notin {\bf E}$ }

Suppose that $ p' \rightarrow q' \in \Sigma$  and $ p' \rightarrow q' $ is
violated in $T$. Then for this to happen the two nodes in $N(q')$ must have
different $val's$  and so by the construction of $T$, $ q'
\rightarrow q \in \Sigma^+$ and so by A3 $ p' \rightarrow q$.  By A4 this
implies that $p \rightarrow q \in \Sigma^+$ which contradicts the
definition of $p_{min1}$ and so we
conclude that either $ p' \rightarrow q'$ is satisfied in $T$ or $ p'
\rightarrow q' \notin \Sigma$   .

\underline{Case ABD.4: $ q' > q$ and $Last(q') \in {\bf E}$ }

Suppose that $p' \rightarrow q' \in \Sigma$. Since $ q' > q$ and $q >
p_{min1}$ then by A6
$q' \rightarrow p_{min1}$ and since $p_{min1} \rightarrow q$ by A3 it follows
that $p' \rightarrow q \in \Sigma^+$. However this contradicts the
definition of
$p_{min1}$ and so $p' \rightarrow q' \notin \Sigma$.

\underline{Case ABD.5: $q' = q_v$, i.e. $ q' > q$ and $Last(q') \notin {\bf
E}$ }

As for ABC.3
\\

\underline{Case ABE: $ p_t > p' $}

As for case ABD.\\

\underline{Case ABF: $ p' = q_w$, i.e. $q > p' \cap q \geq p_{min} $ and
$Last(p') \notin {\bf E}$ }

\underline{Case ABF.1: $q' = q$}

As for case AAK.1.

\underline{Case ABF.2: $q_w > q' \geq p_{min1} $ }

By A6 it follows that $q' \rightarrow p_{min1}$ and since $p_{min1} \rightarrow
q$ it follows by A3 that $q' \rightarrow q$.  Then
following AAK.2,
$ p' \rightarrow q'$ must be satisfied in $T$.

\underline{Case ABF.3: $ q' > q$ and $Last(q') \in {\bf E}$ }

As for case AAG.2 it follows that $q' \rightarrow q$, and so from A3 $p'
\rightarrow q$. So by construction of $T$ the two nodes in $Nodes(p')$ have
different $val's$ and so $p' \rightarrow q'$ is satisfied.

\underline{Case ABF.4: $q' = q_v$, i.e. $ q' > q$ and $Last(q') \notin {\bf
E}$ }

As for AAK.4.\\

\underline{Case ABG: $ p' = q_v$, i.e. $ p' > q \cap p'$ and $Last(p')
\notin {\bf E}$}

\underline{Case ABG.1: $q' = q$}

As for case AAK.1.

\underline{Case ABG.2: $q_w > q' \geq p_{min1} $ }

As for case AAK.2

\underline{Case ABG.3: $ q' > q$ and $Last(q') \in {\bf E}$ }

As for case AAK.3\\

\underline{Case ABH: $p' = p$}

\underline{Case ABH.1: $q_w > q' \geq p_{min1} $ }

Suppose that $p' \rightarrow q' \in \Sigma$. By A6 it follows that $q'
\rightarrow p_{min1}$ and since $p_{min1} \rightarrow
q$ it follows by A3 that $q' \rightarrow q$.  Then
since $p' \rightarrow q'$ applying A3 means that $ p \rightarrow q \in
\Sigma^+$ which is a contradiction and so $p' \rightarrow q' \notin
\Sigma$.

\underline{Case ABH.2: $q' = q_w$, i.e. $q > q' \cap q \geq p_{min}$ and
$Last(q') \notin {\bf E}$  }

Suppose that $ p' \rightarrow q' \in \Sigma$  and $ p' \rightarrow q' $ is
violated in $T$. Then for this to happen the two nodes in $N(q')$ must have
different $val's$  and so by the construction of $T$, $ q'
\rightarrow q \in \Sigma^+$ and so by A3 $ p \rightarrow q$ which is a
contradiction. So we
conclude that either $ p' \rightarrow q'$ is satisfied in $T$ or $ p' \rightarrow q' \notin \Sigma$   .

\underline{Case ABH.3: $ q' > q$ and $Last(q') \in {\bf E}$ }

As for Case ABH.1.

\underline{Case ABH.4: $q' = q_v$, i.e. $ q' > q$ and $Last(q') \notin {\bf
E}$ }

As for Case ABH.2.\\

\underline{Case B: $Last(p) \notin {\bf E}$ }

There are only two cases two consider: (a) $p > q
$; (b) $ p \not> q $ and $q \not>p $.  We consider (b) first.

 \underline{Case BA: $ p \not> q $ and $q \not>p $  }

Let $p_{min}$ and $p_{branch}$ be defined as in Case AA.  We now consider the
two subcases where $p > p_{branch} $ and $ p \not>p_{branch}$.  We now
consider the first case.

 \underline{Case BAA: $p > p_{branch} $ }

Construct then a tree $T$ with the following properties. Firstly $T$ is
complete
w.r.t. $P_\Sigma$.  For all paths $p'$ such that $p' \cap p_{branch}$ is a
strict prefix
of  $p_{branch}$, $T$ contains one path instance for $p'$. If $p' \cap
p_{branch}$
is not a strict prefix of $p_{branch}$ then $T$ contains two path instances
for $p'$
with the following properties.  If $p' = p$ then the $val$ of the two nodes in
$N(p)$ is the same, otherwise if $Last(p') \notin {\bf E}$ then the $val$
of the two
nodes in $N(p')$ are distinct if $p' \rightarrow q \in \Sigma^+$ otherwise
they are
the same.  Such a tree always exists

\begin{figure}[here]
\begin{center}
\includegraphics[scale=.6]{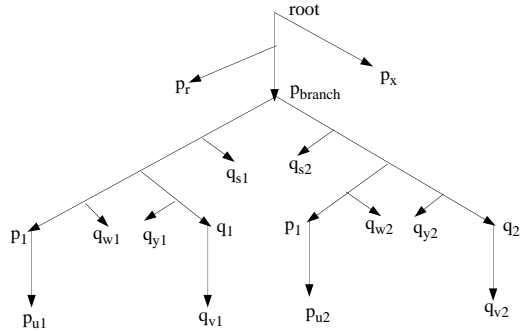}
\end{center}
\caption{A XML tree }
\label{F: Tproof2}
\end{figure}

It is clear from this construction that $T$ violates $p \rightarrow q$ so
it remains
to prove that $T$ satisfies $\Sigma$.  We let $p' \rightarrow q'$ be any
XFD (not necessarily in $\Sigma$).   We shall consider all possible cases
where $p' \rightarrow q'$ may be
violated, and show that either $p' \rightarrow q'$ is satisfied in $T$ or $p'
\rightarrow q'$ cannot be $\Sigma$. There are several cases to consider
depending on where $ p'$ and $q'$
are in the tree.  The different cases can be best illustrated by Figure
~\ref{F: Tproof2}.

\underline{Case BAAA: $p' = p_x$, i.e. $ p' \cap q = root$}

\underline{Case BAAA.1: $q_w > q' \geq p_{min} $ }

As for Case  AAA.1.

\underline{Case BAAA.2: $q' = q_w$, i.e. $q > q' \cap q \geq p_{min}$ and
$Last(q') \notin {\bf E}$  }

As for Case AAA.2.

\underline{Case BAAA.3: $q' = q$}

As for Case AAA.3.

\underline{Case BAAA.4: $ q' > q$ and $Last(q') \in {\bf E}$ }

As for Case AAA.4

\underline{Case BAAA.5: $q' = q_v$, i.e. $ q' > q$ and $Last(q') \notin
{\bf E}$ }

As for Case AAA.2

\underline{Case BAAA.6: $q' = q_s$, i.e. $p \cap q \geq p \cap q' \geq p_{min}$}

Suppose that $p' \rightarrow q' \in \Sigma$ and $p' \rightarrow q'$ is
violated in $T$. Then, by construction of $T$, we
have that
$q_s \rightarrow q \in \Sigma^+$ and so by A3 $p' \rightarrow q$ and so by
A4 $p
\rightarrow q \in \Sigma^+$
which is a contradiction. So $p' \rightarrow q'$ is satisfied or $p'
\rightarrow q' \notin \Sigma$.

\underline{Case BAAA.7: $ q_s  > q' \geq p_{min}$}

Suppose that $p' \rightarrow q' \in \Sigma$. If $ q_s  > q' $ and $q' \geq
p_{min}$ then $q' \rightarrow p_{min}$ by A6
and since $p_{min} \rightarrow q$ then, by A3, $p' \rightarrow q$ which implies
a contradiction as in the previous case and so $p' \rightarrow q' \notin
\Sigma$.

\underline{Case BAAA.8: $q' = q_y$, i.e. $q > q' \geq p \cap q$ }

As for BAAA.6.

\underline{Case BAAA.9: $ q_y  > q' \geq p \cap q $}

As for   AAA.1.

\underline{Case BAAA.10: $q' = p_u$, i.e. $ q' > p$ and $Last(q) \notin
{\bf E}$}

 If $T$ violates $p' \rightarrow q'$, then by construction of $T$ we have
that $q'
\rightarrow q$ and so by A3 $p' \rightarrow q$ which implies a
contradiction as in
case BAAA.6 and so $p' \rightarrow q'$ is satisfied.

\underline{Case BAAA.11: $ p_u  > q' > p$}

Suppose that $p' \rightarrow q' \in \Sigma$. By A6 $q' \rightarrow p_{min}$
and,
since $p_{min} \rightarrow q$ by A3
$p' \rightarrow q$ which implies a contradiction as in BAAA.6 and so $p'
\rightarrow q' \notin \Sigma$.
\\

\underline{Case BAAB: $ p_x  > p' $}

As for Case BAAA.
\\

\underline{Case BAAC: $p' = p_r$, i.e. $ p \cap q \geq p' \cap q > root $}

\underline{Case BAAC.1: $q_w > q' \geq p_{min} $ }

As for case AAC.1

\underline{Case BAAC.2: $q' = q_w$, i.e. $q > q' \cap q \geq p_{min}$ and
$Last(q') \notin {\bf E}$  }

As for case AAC.3

\underline{Case BAAC.3: $q' = q$}

As for case AAC.1.

\underline{Case BAAC.4: $ q' > q$ and $Last(q') \in {\bf E}$ }

As for Case AAC.4

\underline{Case BAAC.5: $q' = v$}

As for Case AAC.5

\underline{Case BAAC.6: $q' = q_s$, i.e. $p \cap q \geq p \cap q' \geq
p_{min}$ and $Last(q') \notin {\bf E}$ }

Suppose $p' \rightarrow q'\in \Sigma$. As in case BAAA.6 we can derive that $p'
\rightarrow q \in \Sigma^+$ which contradicts the
definition of $p_{min}$ and so $p' \rightarrow q' \notin \Sigma$.

\underline{Case BAAC.7: $ q_s  > q' \geq p_{min}$}

Suppose $p' \rightarrow q' \in \Sigma$. As in Case BAAA.7 we can derive
that $p'
\rightarrow q$ which contradicts the
definition of $p_{min}$ and so $p' \rightarrow q' \notin \Sigma$.

\underline{Case BAAC.8: $q' = q_y$, i.e. $q > q' \geq p \cap q$ }

As for BAAC.6.

\underline{Case BAAC.9: $ q_y  > q' \geq p \cap q $ }

As for BAAC.7.

\underline{Case BAAC.10: $q' = p_u$, i.e. $ q' > p$ and $Last(q') \notin
{\bf E}$}

 Suppose $p' \rightarrow q' \in \Sigma$. As in BAAA.10 we can derive that $p'
\rightarrow q$ which which contradicts the
definition of $p_{min}$ and so $p' \rightarrow q' \notin \Sigma$.

\underline{Case BAAC.11: $p_u  > q' > p$}

Suppose $p' \rightarrow q' \in \Sigma$. As in BAAA.11 we can derive that $p'
\rightarrow q$ which which contradicts the
definition of $p_{min}$ and so $p' \rightarrow q' \notin \Sigma$.
\\

\underline{Case BAAD: $ p_r  > p' $}

\underline{Case BAAD.1: $q' = q$}

As for Case AAD.1.
The other cases are similar to the corresponding BAAC cases.
\\

\underline{Case BAAE: $p' = q_s$, i.e. $p \cap q \geq p \cap q' \geq
p_{min}$ and $Last(q') \notin {\bf E}$ }

\underline{Case BAAE.1: $q_w > q' \geq p_{min} $ }

Suppose that $p' \rightarrow q' \in \Sigma$  and $p' \rightarrow q'$ is
violated .  Then since $ q' \geq p_{min}  $ and $Last(q') \in {\bf E}$ then
by A6 $q' \rightarrow p_{min}$  and $p_{min} \rightarrow q$ it follows by
applying A3 twice that $p' \rightarrow
q \in \Sigma^+$.  However, by the construction of $T$ if $p' \rightarrow q \in
\Sigma^+$ then the two nodes in $N(p')$ must have different $val's$.
However, for
$p' \rightarrow q'$ to be violated the two nodes in $N(p')$ must have the
same $val's$ which is a contradiction. So we conclude that  either $p'
\rightarrow q'$ is
satisfied or $p' \rightarrow q' \notin \Sigma$.

\underline{Case BAAE.2: $q' = q_w$, i.e. $q > q' \cap q \geq p_{min}$ and
$Last(q') \notin {\bf E}$  }

Suppose that $p' \rightarrow q' \in \Sigma$ and that $p' \rightarrow q'$ is
violated in $T$.   Then for this to happen the two nodes
in $N(p')$ must have the same $val$ and the two nodes in $N(q')$  must have
different $val's$. So by construction of $T$, $q' \rightarrow q \in
\Sigma^+$ and
hence by A3 this implies that $p' \rightarrow q \in \Sigma^+$.  However, by the
construction of $T$, this implies that the two nodes in $N(p')$ must have
different $val's$ which is a contradiction and so $p' \rightarrow q' \notin
\Sigma$ or $p' \rightarrow q'$ is satisfied in $T$.

\underline{Case BAAE.3: $q' = q$}

If $p' \rightarrow q' \in \Sigma$ then by the construction of $T$ the two nodes
in $N(p')$ have diffferent $val's$ and so $p' \rightarrow q'$ is satisfied.

\underline{Case BAAE.4: $ q' > q$ and $Last(q') \in {\bf E}$ }

Suppose that $p' \rightarrow q' \in \Sigma$. Then by A6 $q' \rightarrow
p_{min}$ and since $p_{min} \rightarrow q$ by A3 we have that $p' \rightarrow
q \in \Sigma^+$.  However, by the construction of $T$, this implies that
the two
nodes in $N(p')$ have diffferent $val's$ and so $p' \rightarrow q'$ is
satisfied.

\underline{Case BAAE.5: $q' = q_v$, i.e. $ q' > q$ and $Last(q') \notin
{\bf E}$ }

Suppose that $p' \rightarrow q' \in \Sigma$. As for case AAC.5 we derive that
$p' \rightarrow q \in \Sigma^+$ and so, as for case BAAE.4, this implies $p'
\rightarrow q'$ is satisfied.

\underline{Case BAAE.6: $q' = q_y$, i.e. $q > q' \geq p \cap q$ }

Suppose that $p' \rightarrow q' $ is violated in $T$.  For this to happen
the two nodes in
$N(q')$ must have different $val's$ and so by the definition of $T$ $q'
\rightarrow
q \in \Sigma^+$.  Then by A3 $p' \rightarrow q \in \Sigma^+$  and so by
definition
of $T$ the two nodes in $N(p')$ must have different $val's$ which
contradicts tha
fact that $p' \rightarrow q' $ is violated and so $p' \rightarrow q' $ is
satisfied.

\underline{Case BAAE.7: $ q_y  > q' \geq p \cap q $}

As for case BAAE.1.

\underline{Case BAAE.8: $ q_s  > q' \geq p_{min}$ }

As for BAAA.7 we derive that $p' \rightarrow q$.  So by the definition of
$T$ the
two nodes in $N(p')$ must have different $val's$  and so $p' \rightarrow
q'$ must
be satisfied.

\underline{Case BAAE.9: $q' = p_u$, i.e. $ q' > p$ and $Last(q') \notin
{\bf E}$}

 As for case BAAE.2.

\underline{Case BAAE.10: $ p_u  > q' > p$}

As for case BAAE.1.
\\

\underline{Case BAAF: $p' = q_w$, i.e. $q > p' \cap q \geq p_{min}$ and
$Last(p') \notin {\bf E}$  }

\underline{Case BAAF.1: $q_w > q' \geq p_{min} $ }

As for case BAAE.1.

\underline{Case BAAF.2: $q' = q_s$, i.e. $p \cap q \geq p \cap q' \geq
p_{min}$ and $Last(q') \notin {\bf E}$ }

As for case BAAE.2.

\underline{Case BAAF.3: $q' = q$}

As for case BAAE.3.

\underline{Case BAAF.4: $ q' > q$ and $Last(q') \in {\bf E}$ }

As for case  BAAE.4.

\underline{Case BAAF.5: $q' = q_v$, i.e. $ q' > q$ and $Last(q') \notin
{\bf E}$ }

As for case BAAE.5.

\underline{Case BAAF.6: $q' = q_y$, i.e. $q > q' \geq p \cap q$ }

As for BAAE.6.

\underline{Case BAAF.7: $ q_y  > q' \geq p \cap q $}

As for case BAAE.7.

\underline{Case BAAF.8: $ q_s  > q' \geq p_{min}$ }

As for BAAE.8.

\underline{Case BAAF.9: $q' = p_u$, i.e. $ q' > p$ and $Last(q') \notin
{\bf E}$}

 As  for case BAAE.2.

\underline{Case BAAF.10: $ p_u  > q' > p$}

As for case BAAA.1
\\

\underline{Case BAAG: $p' = q_y$, i.e. $q > q' \geq p \cap q$ }

As for case BAAF.
\\

\underline{Case BAAH: $p' = p_u$, i.e. $ p' > p$ and $Last(p') \notin {\bf E}$}

\underline{Case BAAH.1: $q_w > q' \geq p_{min} $ }

As for case BAAE.1.

\underline{Case BAAH.2: $q' = q_s$, i.e. $p \cap q \geq p \cap q' \geq
p_{min}$ and $Last(q') \notin {\bf E}$ }

As for case BAAF.2.

\underline{Case BAAH.3: $q' = q$}

As for case BAAF.3.

\underline{Case BAAH.4: $ q' > q$ and $Last(q') \in {\bf E}$ }

As for case  BAAF.4.

\underline{Case BAAH.5: $q' = q_v$, i.e. $ q' > q$ and $Last(q') \notin
{\bf E}$ }

As for case BAAF.5.

\underline{Case BAAH.6: $q' = q_y$, i.e. $q > q' \geq p \cap q$ }

As for BAAF.6.

\underline{Case BAAH.7: $ q_y  > q' \geq p \cap q $}

As for case BAAF.7.

\underline{Case BAAH.8: $ q_s  > q' \geq p_{min}$ }

As for BAAF.8.

\underline{Case BAAH.9: $q' = q_w$, i.e. $q > q' \cap q \geq p_{min}$ and
$Last(q') \notin {\bf E}$  }

 As  for case BAAE.6.

\underline{Case BAAH.10: $ p_u  > q' > p$}

As for case BAAE.1 \\

\underline{Case BAAI: $p' = q_v$, i.e. $ p' > q$ and $Last(p') \notin {\bf E}$ }

As for case BAAH.
\\

\underline{Case BAAJ: $p' = p$}

\underline{Case BAAJ.1: $q_w > q' \geq p_{min} $ }

Suppose $p' \rightarrow q' \in \Sigma$. Following case BAAE.1 we derive
that $p \rightarrow q \in \Sigma^+$ which is a
contradiction. and so $p' \rightarrow q' \notin \Sigma$.

\underline{Case BAAJ.2: $q' = q_w$, i.e. $q > q' \cap q \geq p_{min}$ and
$Last(q') \notin {\bf E}$  }

Suppose that $ p' \rightarrow q' \in \Sigma$  and $ p' \rightarrow q' $ is
violated in $T$. Then for this to happen the two nodes in $N(q')$ must have
different $val's$  and so by the construction of $T$, $ q'
\rightarrow q \in \Sigma^+$ and so by A3 $ p' \rightarrow q$.  By A4 this
implies that $p \rightarrow q \in \Sigma^+$ which is a contradiction and so we
conclude that either $ p' \rightarrow q'$ is satisfied in $T$ or $ p'
\rightarrow q' \notin \Sigma$   .

\underline{Case BAAJ.3: $ q' > q$ and $Last(q') \in {\bf E}$ }

Assume that $ p' \rightarrow q' \in \Sigma$. Then as in case BAAE.4 we derive
the contradiction that $p \rightarrow q \in \Sigma^+$ and so $ p'
\rightarrow q' \notin \Sigma$.

\underline{Case BAAJ.4: $q' = q_v$, i.e. $ q' > q$ and $Last(q') \notin
{\bf E}$ }

As for case BAAJ.2.

\underline{case BAAJ.5: $ q' > q$ and $Last(q') \in {\bf E}$ }

Assume that $p' \rightarrow q' \in \Sigma$. Then as in case BAAE.4 we derive
the contradiction that $p \rightarrow q \in \Sigma^+$ and so $ p'
\rightarrow q' \notin \Sigma$.

\underline{Case BAAJ.6: $q' = q_y$, i.e. $q > q' \geq p \cap q$ }

As for BAAJ.2.

\underline{Case BAAJ.7: $ q_y  > q' \geq p \cap q$}

Assume that $p \rightarrow q' \in \Sigma$. Then as in case BAAE.1 we derive
the contradiction that $p \rightarrow q \in \Sigma^+$ and so $ p'
\rightarrow q' \notin \Sigma$.

\underline{Case BAAJ.8: $q' = q_s$, i.e. $p \cap q \geq p \cap q' \geq
p_{min}$ and $Last(q') \notin {\bf E}$ }

As for case BAAJ.2.

\underline{Case BAAJ.9: $ q_s  > q' $ }

Assume that $p \rightarrow q' \in \Sigma$. Then as in case BAAC.7 we derive
the contradiction that $p \rightarrow q \in \Sigma^+$ and so $ p'
\rightarrow q' \notin \Sigma$.

\underline{Case BAAJ.10: $q' = p_u$, i.e. $ q' > p$ and $Last(q') \notin
{\bf E}$}

As for case BAAJ.2.

\underline{Case BAAJ.11: $ p_u  > q' > p$}

Assume that $p \rightarrow q' \in \Sigma$. Then as in case BAAE.1 we derive
the contradiction that $p \rightarrow q \in \Sigma^+$ and so $ p'
\rightarrow q' \notin \Sigma$.
\\

 \underline{Case BAB: $p \not> p_{branch} $ }

\begin{figure}[here]
\begin{center}
\includegraphics[scale=.6]{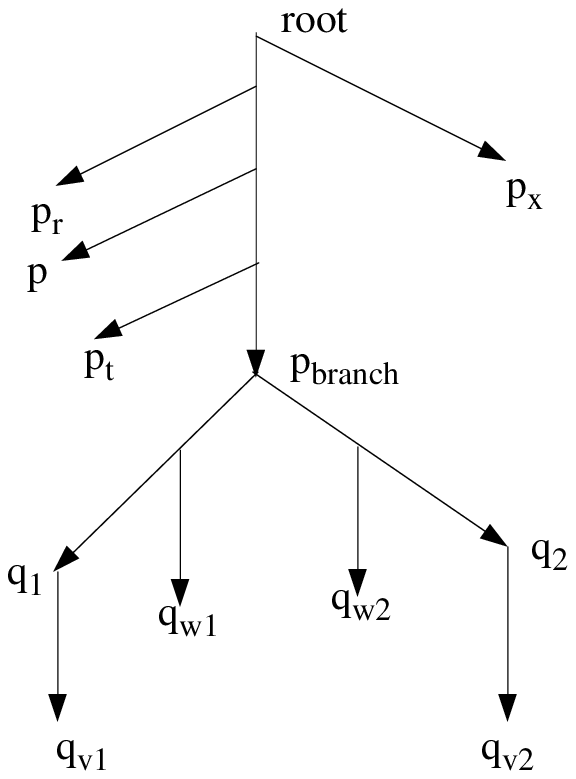}
\end{center}
\caption{A XML tree }
\label{F:Tproof3}
\end{figure}

Construct a tree $T$ as in Case AB. To show that $T$ satisfies $\Sigma$ we let
$p' \rightarrow q'$ be any XFD in $\Sigma$.   We then consider all possible
cases
where $p' \rightarrow q'$ may be violated, and show that either $p' \rightarrow
q'$ is satisfied in $T$ or $p' \rightarrow q'$ cannot be $\Sigma$. The
different cases are illustrated in Figure ~\ref{F:Tproof3}.  Then the same
arguments as in Case AB shows that $T$ satisfies $\Sigma$.

 \underline{Case BB: $p > q $ }

\begin{figure}[here]
\begin{center}
\includegraphics[scale=.6]{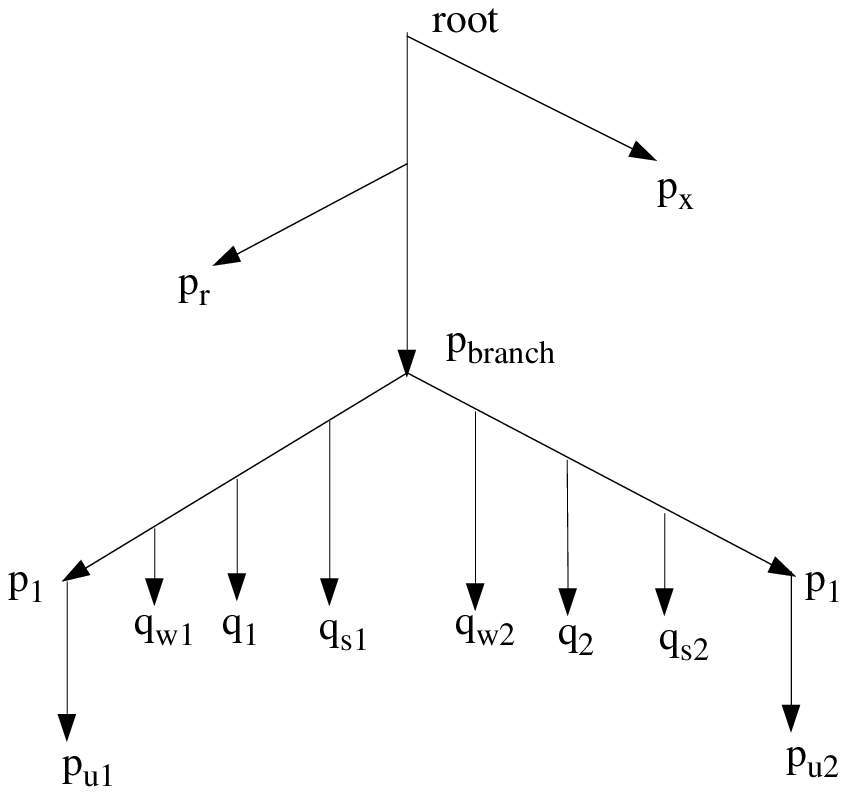}
\end{center}
\caption{A XML tree }
\label{F:Tproof4}
\end{figure}

The first thing we nore is that $q \neq root$ because of A8. Construct then
a tree $T$ as in case BAA. To show that $T$ satisfies $\Sigma$ we let
$p' \rightarrow q'$ be any XFD in $\Sigma$.   We then consider all possible
cases
where $p' \rightarrow q'$ may be violated, and show that either $p' \rightarrow
q'$ is satisfied in $T$ or $p' \rightarrow q'$ cannot be $\Sigma$. The
different cases are illustrated in Figure ~\ref{F:Tproof4}.  Then the same
arguments as in Case BAA shows that $T$ satisfies $\Sigma$.

\hfill$\Box$\\

\end{document}